\documentclass[useAMS,usenatbib]{mn2e}
\usepackage{graphicx,fleqn,natbib,amssymb}
\DeclareGraphicsExtensions{.eps,.cps,.ps,.eps.gz,.ps.gz,.eps.Z}
\usepackage{multirow}
\usepackage{rotating}
\usepackage{dcolumn}

\title[GRB afterglow correlations]{Exploring the canonical behaviour of long gamma-ray bursts using an intrinsic multi-wavelength afterglow correlation.}

\author[Oates et al.]{S. R. Oates$^{1,2}$, J. L. Racusin$^{3}$, M. De Pasquale$^{1,4}$, M. J. Page$^{1}$, A. J. Castro-Tirado$^{2,5}$,
  \newauthor J. Gorosabel$^{2,6,7,\dagger}$, P. J. Smith$^1$, A. A. Breeveld$^1$, N. P. M Kuin$^1$ \\
  $^{1}$ Mullard Space Science Laboratory, University College London, Holmbury St. Mary, Dorking, Surrey, RH5 6NT, UK; sro@iaa.es \\
  $^{2}$ Instituto de Astrofísica de Andaluc\'{i}a (IAA-CSIC), Glorieta de la Astronom\'{i}a s/n, E-18008, Granada, Spain \\
  $^{3}$ Astrophysics Science Division, NASA Goddard Space Flight Center, 8800 Greenbelt Road, Greenbelt, Maryland 20771, USA\\
  $^{4}$ Istituto Astrofisica Spaziale Fisica Cosmica, Palermo, Italy\\
  $^{5}$ Unidad Asociada Departamento de Ingenier\'{i}a de Sistemas y Autom\'{a}tica, E.T.S. de Ingenieros Industriales, Universidad de M\'{a}laga, Spain\\
  $^{6}$ Unidad Asociada Grupo Ciencias Planetarias UPV/EHU-IAA/CSIC, Departamento de F\'{i}sica Aplicada I, E.T.S., \\
  Ingenier\'{i}a, Universidad del Pa\'{i}s Vasco UPV/EHU, Bilbao, Spain\\
  $^{7}$ Ikerbasque, Basque Foundation for Science, Bilbao, Spain \\
  $\dagger$ Deceased}

\begin{document}

\date{Accepted...Received...}

\maketitle

\label{firstpage}

\begin{abstract} 
  In this paper we further investigate the relationship, reported by \cite{oates12}, between the optical/UV afterglow luminosity
  (measured at restframe 200~s) and average afterglow decay rate (measured from restframe 200~s onwards) of long duration Gamma-ray Bursts (GRBs). We extend the analysis
  by examining the X-ray light curves, finding a consistent correlation. We therefore explore how the parameters of these correlations relate to the prompt emission phase
  and, using a Monte Carlo simulation, explore whether these correlations are consistent with predictions of the standard afterglow model. We find significant
  correlations between: $\rm log\;L_{O,200\rm{s}}$ and $\rm log\;L_{X,200\rm{s}}$; $\alpha_{O,>200\rm{s}}$ and $\alpha_{X,>200\rm{s}}$, consistent with simulations.
  The model also predicts relationships between $\rm log\;E_{iso}$ and $\rm log\;L_{200,\rm{s}}$, however, while we find such relationships
  in the observed sample, the slope of the linear regression is shallower than that simulated and inconsistent at $\gtrsim 3\sigma$. Simulations also do not
  agree with correlations observed between $\rm log\;L_{200\rm{s}}$ and $\alpha_{>200\rm{s}}$, or $\rm log\;E_{iso}$ and $\alpha_{>200\rm{s}}$. Overall, these
  observed correlations are consistent with a common underlying physical mechanism producing GRBs and their afterglows regardless of their detailed
  temporal behaviour. However, a basic afterglow model has difficulty explaining all the observed correlations. This leads us to briefly discuss
  alternative more complex models.
\end{abstract}


\section{Introduction}
\label{intro}
Gamma-ray bursts (GRBs) are intense flashes of gamma-rays that are usually accompanied by an afterglow, longer lived emission that 
may be detected at X-ray to radio wavelengths. Studies of single GRBs provide exceptional detail on the behaviour and physical 
properties of individual events. However, statistical investigations of large samples of GRBs aim to find common characteristics 
and correlations that link individual events and therefore provide insight into the mechanisms common to GRBs. Statistical 
investigations performed so far have found a number of trends and correlations within and linking the prompt gamma-ray emission and the 
afterglow emission \citep[e.g.,][]{ama02,ghi04,dai08,pan08,berna12,dav12,li12,lia13,zan13,pan13}. Within the prompt gamma-ray emission, 
the most renowned correlation discovered is the Amati relation, a correlation between the isotropic $\gamma$-ray energy $\rm E_{iso}$ and
the restframe $\gamma$-ray peak energy $\rm E_{peak}$ \citep[see][and references therein]{ama02}. The exact origin of the correlation is
uncertain, but it can be explained by the non-thermal synchrotron model, jets viewed over a range of viewing angles or with jets of 
different non-uniform structure \citep{ama06} and can also be produced in the photospheric model \citep{laz13}. Within the afterglow,
several trends are apparent and are currently being explored. The simplest observation is that the luminosity light curves, in
both the X-ray and optical/UV samples, show clustered behaviour. Evidence for clustering into two groups, at $\sim0.5-1$ day, for
the optical/IR GRB afterglows was reported by \cite{boe00,gen05,nard06,liang,nard08, gen08b}. However, several more recently published
works have suggested that the afterglow distributions are unimodal \citep{mel08,cenko09,oates09,kan10,mel14}.

There has also been a suggestion that in samples of pre-{\it Swift} X-ray afterglow light curves with bimodal luminosity distributions,
the more luminous cluster decays more quickly than the less luminous cluster of X-ray afterglows \citep{boe00,gen05}, implying a possible
relationship between the brightness of the GRB afterglow and the rate that it decays. \cite{gen08} extended their X-ray afterglow sample
to include the first 1.5 years of {\it Swift} light curves. Using data from the end of the X-ray plateau phase (between 200~s and 129~ks)
onwards, they again observed clustering into two groups (three groups including low-luminosity GRBs), but they could not support previous
claims that the brighter cluster of GRBs decay typically faster than the fainter cluster of GRBs. 

A relationship between the intrinsic brightness and rate of decay of GRBs has also been explored in other studies. \cite{kou04} explored
a sample of 15 X-ray luminosity light curves from a mix of GRBs and supernovae (SNe). With extrapolation of the GRB X-ray light curves
to a few thousand days after the trigger, the initially broad luminosity distribution of GRB and SNe light curves narrows with time by
an order of magnitude, suggesting the brightest decayed more quickly. In \cite{oates09}, within a sample of 27 {\it Swift}
Ultra-violet Optical Telescope \citep[UVOT;][]{roming} afterglow light curves, observed between $<400$~s and $>10^5$~s, a correlation
was noticed in the observed frame between the magnitude of the $v$-band afterglow light curve at 400~s and the average rate
at which the light curves decayed. A restframe correlation proved inconclusive due to the small sample size. The cluster of luminosity
light curves in \cite{kan10}, showed evidence for narrowing of the distribution with time, also suggesting a relationship between the
brightness of the afterglow and the rate of decay. In \cite{oates12}, the UVOT sample was extended to 48 optical/UV GRB light curves.
Consistent with \cite{kan10} and other studies mention above, the optical/UV luminosity light curves clustered into a single group
and it was apparent that the luminosity distribution was wider during the early part of the afterglow, and became narrower as the
afterglows faded. This finding suggests that the most luminous GRB afterglows at early epochs, decay more quickly than the less luminous
afterglows. Using the logarithmic optical brightness ($\rm log\;L_{\rm O,200s}$; measured at restframe 200~s and at a restframe wavelength 1600${\rm\AA}$),
 and average decay rate of GRB afterglows ($\alpha_{\rm >200s}$; measured from restframe 200~s onwards with a single power-law and
thus ignoring the precise temporal behaviour of the afterglow), \cite{oates12} tested to see if this correlation was statistically significant. With
a Spearman rank test a coefficient of $-0.58$ at a significance of 99.998 per cent (4.2$\sigma$) was found, indicating that these two parameters
are correlated. This correlation is interesting since it does not depend on detecting certain temporal features and is independent of
the shape of the light curves and therefore applicable to essentially all long GRB afterglows.

In this paper, we use the same sample as \cite{oates12} to further explore the $\rm log\;L_{\rm 200s}-\alpha_{\rm >200s}$ relation observed in
the optical/UV. We wish to examine whether this correlation is observed also in the X-ray and how it relates to other GRB properties. Since the
observed X-ray-optical emission is predicted by the standard afterglow synchrotron model, currently the favoured scenario in terms of
producing the afterglow, we will begin by predicting the relationships we should expect to observe for a sample of 48 GRBs. In this model, there
is typically more than one equation to describe the relationship between two parameters. The precise equation depends on the circumstellar environment
and spectral regime, which will be different for different GRBs. Therefore we use a Monte Carlo simulation to predict the expected overall relationship
for a group of GRBs with similar parameters to our sample. We also extend our analysis to include comparisons between the afterglow parameters,
$\rm log\;L_{\rm 200s}$ and $\alpha_{\rm >200s}$, with prompt emission phase parameters, namely the isotropic energy $E_{iso}$, the peak energy
$E_{peak}$ and the duration over which 90 per cent of the prompt emission was observed $\rm T_{90}$. 

This paper is organized as follows. We define our sample in \S~\ref{reduction} and in \S~\ref{linear} we discuss the linear regression methods
we shall use throughout the paper. In \S~\ref{theory}, we present the analytic correlations expected from the standard afterglow model and in \S~\ref{MC}
we present the correlations predicted by the Monte Carlo simulation. In \S~\ref{results} we look at whether we observe these correlations within 
our sample of X-ray and optical/UV luminosity light curves and compare the findings with the relationships predicted by the standard afterglow model
in \S~\ref{discussion}. Finally we conclude in \S~\ref{conclusions}. All uncertainties throughout this paper are quoted at
1$\sigma$. The temporal and spectral indices, $\alpha$ and $\beta$, are given by the expression $F(t,\nu)\propto t^{\alpha}\nu^{\beta}$.
Throughout, we assume the Hubble parameter $H_0 = 70\;{\rm km s}^{-1}\;{\rm Mpc}^{-1}$ and density parameters $\Omega_{\Lambda}= 0.7$ and $\Omega_m= 0.3$. 

\section{GRB afterglow sample}
\label{reduction}
Our sample contains the same GRBs examined in \cite{oates12}. The sample consists of 56 long duration GRBs
with optical/UV afterglows, selected from the second {\it Swift} UVOT GRB afterglow catalogue \citep{rom}, which were observed
between April 2005 and December 2010. They were selected using the criteria of \cite{oates09}: the optical/UV light curves must
have a peak UVOT $v$-band magnitude of $\leq$17.89 (equivalent to a count rate of 1 $\rm s^{-1}$), UVOT must observe within 
the first 400~s until at least $10^5$~s after the BAT trigger and the colour of the afterglows must not evolve significantly with 
time, meaning that at no stage should the light curve from a single filter significantly deviate from any other filter light curve 
when normalized to the $v$ filter. These criteria ensure that a high signal-to-noise (SN) light curve, covering both early and 
late times, could be constructed from the UVOT multi-filter observations \citep[see][for further details]{oates09,oates12}. 
Furthermore, these GRBs have spectroscopic or photometric redshifts and we were able to determine the host E(B-V) 
values \citep[the host extinction was derived from spectral energy distributions constructed from the afterglow emission 
following the methodology in][]{sch10}. For each GRB, optical luminosity light curves were produced at a common wavelength of 1600~{\AA} \citep{oates12}. 
This wavelength was selected to maximise the number of GRBs with spectral energy distributions that covered this 
wavelength and to be relatively unaffected by host extinction. A k-correction factor, $k$, was computed for each GRB. This was 
taken as the flux density at the wavelength that corresponds to 1600~{\AA} in the rest frame, $F_{1600}$, divided by 
the flux density at the observed central wavelength of the $v$ filter (5402~{\AA}), $F_{v}$, which was multiplied by 
$(1+z)$, where $z$ is the redshift of the GRB such that $k=(F_{1600}/(F_{v}*(1+z)))$. For those GRBs with SEDs not 
covering 1600~{\AA}, an average $k$ value was determined from the other GRBs in the sample, which have SEDs covering 
both 1600~{\AA} and the $v$ filter rest frame wavelength. The time of each light curve was corrected to the restframe 
by $t_{rest}=t_{obs}/(1+z)$. The luminosity light curves were also corrected for Galactic and host extinction. 

All 56 GRBs in the optical/UV sample have X-ray counterparts. The X-ray light curves were retrieved from
the University of Leicester {\it Swift} XRT GRB data repository \citep{eva07,eva09}. The $0.3-10$\,keV flux light
curves were converted to luminosity at restframe $1$\,keV. They were k-corrected using a k-correction of $(1+z)^{-(1+\beta)}$ \citep[e.g][]{ber03b},
where the $\beta$ is from spectral modeling. The time of each light curve was corrected to the restframe by $t_{rest}=t_{obs}/(1+z)$.
The X-ray luminosity light curves were also corrected for Galactic and host neutral hydrogen absorption. 

We selected restframe 200s as the time to obtain the luminosity and the time from which to fit a power-law to the
afterglow light curves since before this time the optical afterglows are variable and may be rising to a peak. This
behaviour typically ends before restframe 200s. Also by this time, the initial steep decay segment for the
majority of X-ray light curves in our sample with this feature ceases. This steep decay segment is likely the
tail of the prompt emission \citep{zhang06}. Therefore for each GRB, we interpolated the optical luminosity at
200~s using data between 100 and 2000s and for the X-ray we measured the luminosity at 200~s from the best fit
light curve model \citep{racusin09}. To obtain the average decay rate, we fit a single power-law to each optical
and X-ray light curve using data from 200~s onwards. For 8 optical/UV light curves, we were unable to determine
one or both of the luminosity at 200~s and the average decay index. We therefore excluded these GRBs from our sample.

While the initial steep decay is not observed at restframe 200~s for most of the X-ray light curves in our sample,
it is present at restframe 200~s for 8 GRBs. We identify a light curve segment to have a prompt origin if there
is a steep to shallow transition with $\Delta\alpha>1.0$. In these situations the average decay index is
measured with a simple power-law fit to data beyond restframe 200~s and after the steep to shallow transition. In
order to get a better estimate of the afterglow luminosity at restframe 200~s, we extrapolate back to restframe 200~s
the first segment of the best fit light curve that is not contaminated by the prompt emission \citep[see also][]{racusin15}.

We do not observe flares in the optical/UV or X-ray light curves at restframe 200s. However, flares present after restframe 200s
may affect in the observed average decay index and therefore introduce some scatter in correlations. Furthermore, the X-ray
shallow decay segment may be comprised of emission from the prompt and afterglow phases. An indication of this would be
evolution of the X-ray hardness ratio as the light curve transitions from being a combination of the prompt and afterglow
emission to only produced by the afterglow. We have checked the X-ray hardness ratios, from restframe 200~s onwards, for all
the X-ray afterglows in our sample and we do not find strong evidence for evolution, suggesting that the prompt emission does not
strongly affect the X-ray afterglow for these GRBs. A reverse shock is also expected to be observed in the early optical/UV light curve,
but is not commonly observed \citep{oates09}. We can assume that at restframe 200~s, the reverse shock has either ceased or
contributes at a similar or lower level as the forward shock emission for the optical/UV light curves in our sample. However,
a reverse shock could also be a cause of scatter in the correlations involving parameters from the optical/UV afterglow.

In order to compare the afterglow properties with the prompt emission properties we determined the isotropic $\gamma$-ray 
energy $\rm E_{\rm iso}$ and peak energy, $\rm E_{\rm peak}$ from the $\gamma$-ray emission, following \cite{racusin09}.
The BAT fluence was converted to $\rm E_{\rm iso}$ at a rest-frame bandpass of 10 to 10000~keV using equation 4 from \cite{blo01}.
The k-correction was computed using the Band Function \citep{band93}. Where available, the 
spectral parameters for the Band function were obtained from the 2nd {\it Swift} BAT catalogue \citep{sak11}, the {\it Fermi} 
GBM catalog \citep{pac12} and {\it Konus-Wind} GCNs. When available, we used the measured spectral slopes, otherwise we assume 
$\alpha = −1$ and $\beta = −2.5$. In the cases where no $\rm E_{\rm peak}$ was reported we used the correlation between the 
peak energy and the photon index of the $\nu F_\nu$ spectrum to estimate $\rm E_{\rm peak}$ \citep[see][for further details]{sak09}. 
The relationship can only be used to estimate $\rm E_{\rm peak}$ when the power-law index of the BAT spectrum is between -2.3 and 
-1.3, which places $\rm E_{\rm peak}$ approximately within the BAT range \citep[see also][for further details]{racusin09}.
For 3 GRBs, $\rm E_{\rm peak}$ was not reported and we were unable to use the Sakamoto relation to provide an estimate. In these cases, when
calculating  $\rm E_{\rm iso}$ we assumed a power-law spectrum. Of the 48 GRBs in our sample, we were able to determine
$\rm E_{\rm peak}$ for 44 and $\rm E_{\rm iso}$ for 47 GRBs. Furthermore, it is difficult to reliably determine the errors on
$\rm E_{\rm peak}$ and $\rm E_{\rm iso}$ and so we only have error bars for a handful of them. As detailed in the
next section, when performing the linear regression involving $\rm E_{\rm peak}$ or $\rm E_{\rm iso}$, we did not use the
{\textsc FITEXY} IDL regression routine, but rather {\textsc SIXLIN} IDL code, which does not require errors on either parameter.
However, by using SIXLIN we are assuming each point has similar weighting. This may not be the case since $\rm E_{\rm peak}$ is
derived from two different methods. The Sakamoto relationship is an estimate of the likely $\rm E_{\rm peak}$ and typically
has a $1\sigma$ uncertainty in $\rm E_{\rm peak}$ that is larger than the 90\% error found for BAT derived $\rm E_{\rm peak}$
values. All the main parameters used in this paper for the correlations can be found in the appendix, see Table \ref{appenx1}.

\section{Linear Regression}
\label{linear}
For the linear regression we use the IDL routines {\textsc FITEXY} and {\textsc SIXLIN}: {\textsc FITEXY} is used when both
parameters have errors, {\textsc SIXLIN} is used when we do not know the errors on one or both parameters. Since there are
only a handful of GRBs with errors on the $\rm E_{\rm iso}$ and $\rm E_{\rm peak}$ parameters in order to
maintain a large number of events for the regression, we choose to discard errors in both parameters. We therefore use the
{\textsc SIXLIN} regression routine when one of the parameters involved is $\rm E_{\rm iso}$, $\rm E_{\rm peak}$ or $\rm{T90}$.

The routine {\textsc SIXLIN} produces the results of 6 linear regression methods outlined in \cite{iso90}. 
We want to determine the best physical relationship between two parameters, not the predictive 
relationship that results in a value of $y$ given $x$, which is typically irreversible (i.e. for 
linear regressions $y=m_1x+c_1$ \& $x=m_2y+c_2$, $m_1\neq 1/m_2$ and $c_1\neq c_2/m_2$). Of the six routines in the {\textsc SIXLIN} function, \cite{iso90} 
recommend the bisector model, which is independent of the choice of $x$ and $y$. 
This routine determines the mean slope between the ordinary least squares regression of $x$ versus $y$, and $y$ versus $x$. However, when there 
is little or no correlation between the parameters, the resulting bisector slope can be mis-leading\footnote{For 
instance if there is little correlation between two parameters, with points spread out along the $x$-axis, the ordinary least squares 
regression of $x$ versus $y$ would give a slope close to zero, while $y$ versus $x$ would give a large value. The bisector 
regression model would return a slope somewhere in between.} if not read in conjunction with the Spearman 
rank coefficient. This is a particular issue when performing the Monte Carlo simulation in \S~\ref{discussion}. We 
therefore, report the best-fit regression using the orthogonal regression. The orthogonal regression method is symmetric, providing a
consistent result regardless of whether the regression is applied to $x$ versus $y$ or $y$ versus $x$. This method is only
recommended if the parameters involved are scale-free, i.e. logarithmic, or are scale-invariant \citep{iso90}. In this paper,
it is appropriate to use the orthogonal regression since the temporal decay index is scale invariant and all other parameters
are ratios or are logarithms.

A final point raised by \cite{iso90} \& \cite{fei92}, which we address at the end of this section, is that for small samples (N$<50$) 
the errors on the regression parameters may be underestimated and in this case it is more appropriate to use a bootstrap method to 
provide an estimate of these errors.

The routine {\textsc FITEXY} is based on the procedure provided by \cite{press92} and also has the advantage that the input variables 
$x$ and $y$ are treated symmetrically so we do not need to assume that $x$ is the independent variable and $y$ is the dependent 
variable. However, while this method takes into account measurement errors in both parameters, it does not take in to account 
intrinsic scatter in the data. The estimates of the errors of the slope and constant parameters are therefore typically too small 
and again in this case it is more appropriate to use a bootstrap method to provide an estimate to the errors on the 
regression parameters.

We therefore chose to determine the errors for both routines using the bootstrap method. For $10^4$ trials, we 
randomly selected from the input data, a sample of points the same size as the input data. After one point was selected at 
random, we returned it to the set of observed data points, allowing it to be selected more than once during each trial. Once 
a set of points had been selected equal in size to the observed data set, we ran {\textsc FITEXY} or {\textsc SIXLIN} on this
set of points. For each of the $10^4$ trials, we recorded the slope and constant value. To provide the 1$\sigma$ errors,
we separately ordered the recorded sets of slope and constant values by size and selected the upper and lower errors as the
difference between the mean and the values at 15.9 per cent and 84.1 per cent. During this process a Spearman rank
correlation was also performed on the simulated data so that we could obtain the $1\sigma$ errors given in
Table \ref{Spear_lin} in a similar fashion.

\section{The standard synchrotron afterglow model}
\label{theory}
The standard afterglow synchrotron model is currently the favoured scenario in terms of producing the observed X-ray-optical afterglow emission. 
In this model, the afterglow is a natural result of the collimated ejecta reaching the external medium and interacting with it, producing the
observed synchrotron emission. For a given frequency, the observed flux depends on the position of the frequency relative to the
synchrotron frequencies (the synchrotron cooling frequency $\nu_c$, the synchrotron peak frequency $\nu_m$ and the synchrotron self-absorption
frequency $\nu_a$) and the values of the microphysical parameters (the kinetic energy of the outflow $E_k$, the fraction of energy given to the
electrons $\epsilon_e$, the fraction of energy given to the magnetic field $\epsilon_B$, the structure and density of the external medium and the
electron energy index $p$). Therefore it is possible to predict the relationships between observable and/or microphysical parameters at any time
during the afterglow \citep[e.g.,][]{sar98,pan00,gao13}. Since the kinetic energy $E_{k}$ and the isotropic energy $E_{iso}$ are related linearly through
the efficiency parameter $\eta=E_{iso}/(E_{iso}+E_{k})$ and the luminosity is a function of $E_{k}$, we can also predict the relationship between
optical/UV and X-ray luminosities and $E_{iso}$.

We now derive the relationships we should expect in our observed sample of optical/UV and X-ray luminosity light curves. We use the expectations
for flux in \cite{sar98} (see their eq 8.) and the equations for peak flux, $F_{\nu,max}$, $\nu_c$ and $\nu_m$ in \cite{zhang07}. We assume an
isotropic, collimated outflow which is not energy injected and since we wish to consider a very simplistic model we do not consider the emission
from the traditional reverse shock \cite[e.g.,][]{zhang03}. We can justify excluding the contribution from the reverse shock because we only
examine parameters at or beyond restframe 200~s, by which time the reverse shock in most cases has either ceased or contributes at a similar
or lower level to the forward shock emission \citep[e.g.,][]{oates07}.

Studies of individual GRBs and samples of GRBs suggest that a large fraction of afterglows are produced by outflows ploughing into constant density
media \citep[e.g.][]{ryk09,oates09,sch11}. Therefore we shall only consider relationships appropriate for this density medium. This assumption
will be verified later in \S~\ref{discussion}. As the optical/UV and X-ray emission is likely to be above $\nu_m$ at restframe 200~s, we also
only consider the most likely spectral regimes either $\nu_m<\nu_{O}<\nu_c<\nu_{X}$,\quad$\nu_m<\nu_{O}<\nu_{X}<\nu_c$ or $\nu_m<\nu_c<\nu_{O}<\nu_{X}$. 

Under these conditions there are three possible relationships between optical and X-ray luminosity in the standard afterglow model:

\begin{equation}
  L_X=\left\{
  \begin{array}{lr}
    \left(\frac{\nu_X}{\nu_O}\right)^{-(p-1)/2}L_O & : \nu_m<\nu_O<\nu_X<\nu_c \\
    \left(\frac{\nu_X}{\nu_c}\right)^{p/2}\left(\frac{\nu_O}{\nu_c}\right)^{-1/2}L_O & : \nu_m<\nu_O<\nu_c<\nu_X \\
    \left(\frac{\nu_X}{\nu_O}\right)^{-p/2}L_O & : \nu_m<\nu_c<\nu_O<\nu_X 
  \end{array}
  \right.
  \label{lum_eq}
\end{equation}

where $p$ is the energy distribution index ($dN(E)/dE\propto E^{p}$; where N(E) is the number of electrons with energy, E). Each scenario
predicts a linear relationship, with a normalization that is dependent on $p$, and for the second regime only, also dependent on the 
value of $\nu_c$. These relationships suggest that for a distribution of $\rm L_{X}$ versus $\rm L_{O}$, we may expect to observe two
different lines, corresponding to the first and third relationships of Eq. \ref{lum_eq}, bridged together by data points corresponding
to the $\nu_m<\nu_O<\nu_c<\nu_X$ relationship.

The standard afterglow model also predicts relationships between $\rm L_{\nu}$, at a given frequency $\nu$, with kinetic energy $E_{k}$ as: 

\begin{equation}
  L_\nu \propto \left\{
  \begin{array}{lr}
    E_{k}^{\left(\frac{p+3}{4}\right)} & : \nu_m<\nu<\nu_c \\
    E_{k}^{\left(\frac{p+2}{4}\right)} & : \nu_m<\nu_c<\nu 
  \end{array}
  \right.
  \label{energy_eq}
\end{equation}
Predicting the observed relationship is complicated by the fact that we need to know the value for the efficiency in order to get the
direct relationship between $L_{200\rm{s}}$ and $E_{iso}$.

Finally, we can easily show what we should expect, in terms of the standard afterglow model, for the relationship of
$(L_{O}/L_{X})$ with energy:

\begin{equation}
  \frac{L_O}{L_X}= \left\{\hspace{-0.2cm}
  \begin{array}{lr}
    \left(\frac{\nu_O}{\nu_X}\right)^{-(p-1)/2}  & \hspace{-0.25cm}: \nu_m<\nu_O<\nu_X<\nu_c \\
    \left(\frac{\nu_O}{\nu_X}\right)^{-p/2}\hspace{-0.1cm}\left(\frac{\nu_{O}}{\nu_c}\right)^{1/2} \hspace{-0.1cm}\propto E_{k}^{1/4} & \hspace{-0.25cm}: \nu_m<\nu_O<\nu_c<\nu_X \\
    \left(\frac{\nu_O}{\nu_X}\right)^{-p/2} & \hspace{-0.25cm}: \nu_m<\nu_c<\nu_O<\nu_X 
  \end{array}
  \right.
  \label{rat_energy_eq}
\end{equation}

When the optical/UV and X-ray bands lie on the same segment, $\rm log\;(L_{O}/L_{X})$ is independent of 
$\rm log\;{E_{iso}}$, but is dependent on $p$. Assuming a range for $p$ of between 2.0 and 3.0, this ratio of the optical/UV and X-ray luminosities lies 
between 1.05 and 3.16 (where $\nu_O=1.87\times10^{15}$~Hz and $\nu_{X}=2.4\times10^{17}$~Hz). When the X-ray and optical/UV 
bands lie on different segments, the ratio will range between 1.05 and 3.16, but the ratio is dependent on the choice of energy, 
such that the ratio increases with $E_{k}^{1/4}$. 

In Eqs. \ref{lum_eq},\ref{energy_eq} \& \ref{rat_energy_eq}, there are several possibilities for how the parameters are related and it 
is likely that different GRBs are in different regimes and so satisfy different formulae. This makes a simple analytic 
prediction of the expected relationships in a sample of observed parameters difficult to determine. Therefore in \S~\ref{MC} we 
use a Monte Carlo simulation to predict the correlations we should actually observe, between these and other parameters, when 
using a sample of GRB afterglows.

\section{Monte Carlo Simulation}
\label{MC}
Since the standard afterglow model does not offer a single equation for the relationships between parameters, we employed 
the use of a Monte Carlo simulation, to determine what relationships we should expect from this model with the 
same number of GRBs as our sample. Using $10^4$ trials, we simulated the optical/UV (at 1600~\AA) and X-ray (at 1~keV) flux densities for 
48 GRBs using equation 8 of \cite{sar98} and equations 4, 5 and 6 given in \cite{zhang07} for
$F_{\nu,max}$, $\nu_m$ and $\nu_c$. In this simulation we assume that all GRBs are produced in a constant density medium, consistent
with our assumption detailed in \S \ref{theory}. To compute $F_{\nu,max}$, $\nu_m$ and $\nu_c$ we needed to provide values for the 
microphysical parameters. These were selected at random from log-normal distributions which had 3$\sigma$ intervals 
ranging between: 0.01-0.3 for the fraction of energy given to the electrons, $\epsilon_e$; $5\times10^{-4}-0.5$ for the fraction 
of energy given to the magnetic field, $\epsilon_B$, and $10^{-3}-10^{3}{\rm cm^{-2}}$ for the density of the external medium. The centre of each 
of these distributions is at the logarithmic midpoint. For the electron energy index $p$, we centred the distribution at 2.4, as determined 
by \cite{cur09}, however, we set the 1$\sigma$ width to be 0.2 rather than 0.59. Since the closure relations fail for $p$ values $<2$, 
we re-sampled the $p$ value when $p<2$ was selected. The value of $p$ along with the position of $\nu_c$ relative to the observed 
band and redshift (selected from a uniform distribution with the range 0.5 - 4.5, a similar range as the observed sample), 
dictate the values of $\alpha$, $\beta$ and the k-correction \citep[as given in][]{ber03b}. 

For the 48 GRBs in each trial, we selected a prompt emission energy from a log-normal distribution with a $3\sigma$ range $10^{51}-10^{54}$ erg. 
This range and distribution was selected to be similar to that of the GRBs in this paper, e.g., Table \ref{appenx1}. We picked a random 
value between 10 per cent and 99 per cent for the efficiency, which we used to convert the prompt 
emission energy into kinetic energy. Once all the microphysical parameters, redshift and kinetic energy had been selected, we 
were then able to determine the position of $\nu_c$ and thus knew where it was in relation to $\nu_O$ and $\nu_X$. With this 
information, we then calculated the value of the optical and X-ray fluxes and converted these to luminosity; a k-correction 
was applied during this conversion. We finally took the logarithm of both parameters. As a byproduct of calculating 
the optical and X-ray luminosities, we also have simulated distributions for $\rm E_{iso}$ and $\alpha$. Therefore we 
also produce predictions for comparisons that involve these parameters in addition to those examined in \S~\ref{theory}. 

Once a sample of 48 GRBs had been constructed, we then performed linear regressions, using the IDL routine SIXLIN, and we also 
calculated the Spearman rank coefficient. We repeated the above until we completed all $10^4$ trials. From this routine we 
obtained the best fit slopes to the correlations between several parameters: the optical/UV and X-ray luminosities, the 
optical/UV and X-ray decay indices and $\rm E_{iso}$. The pairs of parameters can be found in Table \ref{MCsim}. For each 
distribution of slope values we take the mean and the 1$\sigma$ error to be the difference between the mean and the 15.9 per 
cent and 84.1 per cent values (when the data are ordered numerically).

\begin{table*}
\centering
\begin{tabular}{llD{,}{\pm}{1.2}D{,}{\pm}{-1.2}D{,}{\pm}{1.2}}
\hline
\multicolumn{2}{c}{Parameters} & \multicolumn{1}{c}{Simulated Spearman} & \multicolumn{2}{c}{---Best fit linear regression for simulation---} \\
\multicolumn{1}{c}{$x$-axis} & \multicolumn{1}{c}{$y$-axis} & \multicolumn{1}{c}{Rank Coefficient} & \multicolumn{1}{c}{Slope} & \multicolumn{1}{c}{Constant} \\
\hline
\hline
$\rm log\;L_{O,200\rm{s}}$ & $\rm log\;L_{X,200\rm{s}}$ & 0.92,0.02  & 0.82,0.04   & 3.76,1.25   \\
$\alpha_{O,>200\rm{s}}$    & $\alpha_{X,>200\rm{s}}$    & 0.74,0.06  & 1.10,0.15   & -0.04,0.17   \\[0.3cm]

$\rm log\;L_{O,200\rm{s}}$ & $\alpha_{O,>200\rm{s}}$    & -0.30,0.14 & -0.04,0.02  & 0.31,0.65    \\
$\rm log\;L_{X,200\rm{s}}$ & $\alpha_{X,>200\rm{s}}$    & -0.20,0.14 & -0.04,0.03  & -0.10,0.78    \\[0.3cm]

$\rm {log\;E_{iso}}$ & $\alpha_{O,>200\rm{s}}$         & -0.06,0.15 & -0.03,0.06  & 0.32,2.91  \\
$\rm {log\;E_{iso}}$ & $\alpha_{X,>200\rm{s}}$         & -0.09,0.15 & -0.04,0.06  & 0.76,3.13  \\
$\rm {log\;E_{iso}}$ & $\rm log\;L_{O,200\rm{s}}$      & 0.51,0.11  & 4.43,1.03   & -200.76,54.10 \\
$\rm {log\;E_{iso}}$ & $\rm log\;L_{X,200\rm{s}}$      & 0.54,0.11  & 3.28,0.71   & -142.22,37.33  \\%[0.3cm]

\hline
\end{tabular}
\caption{The Spearman rank coefficient and linear regression parameters as predicted by the synchrotron model for a sample of 48 GRBs. 
These values were computed with a Monte Carlo simulation with $10^4$ trials.}
\label{MCsim}
\end{table*}

\begin{figure}
\includegraphics[trim=0cm 0cm 0cm 5.5cm,clip=true,angle=0,scale=0.45]{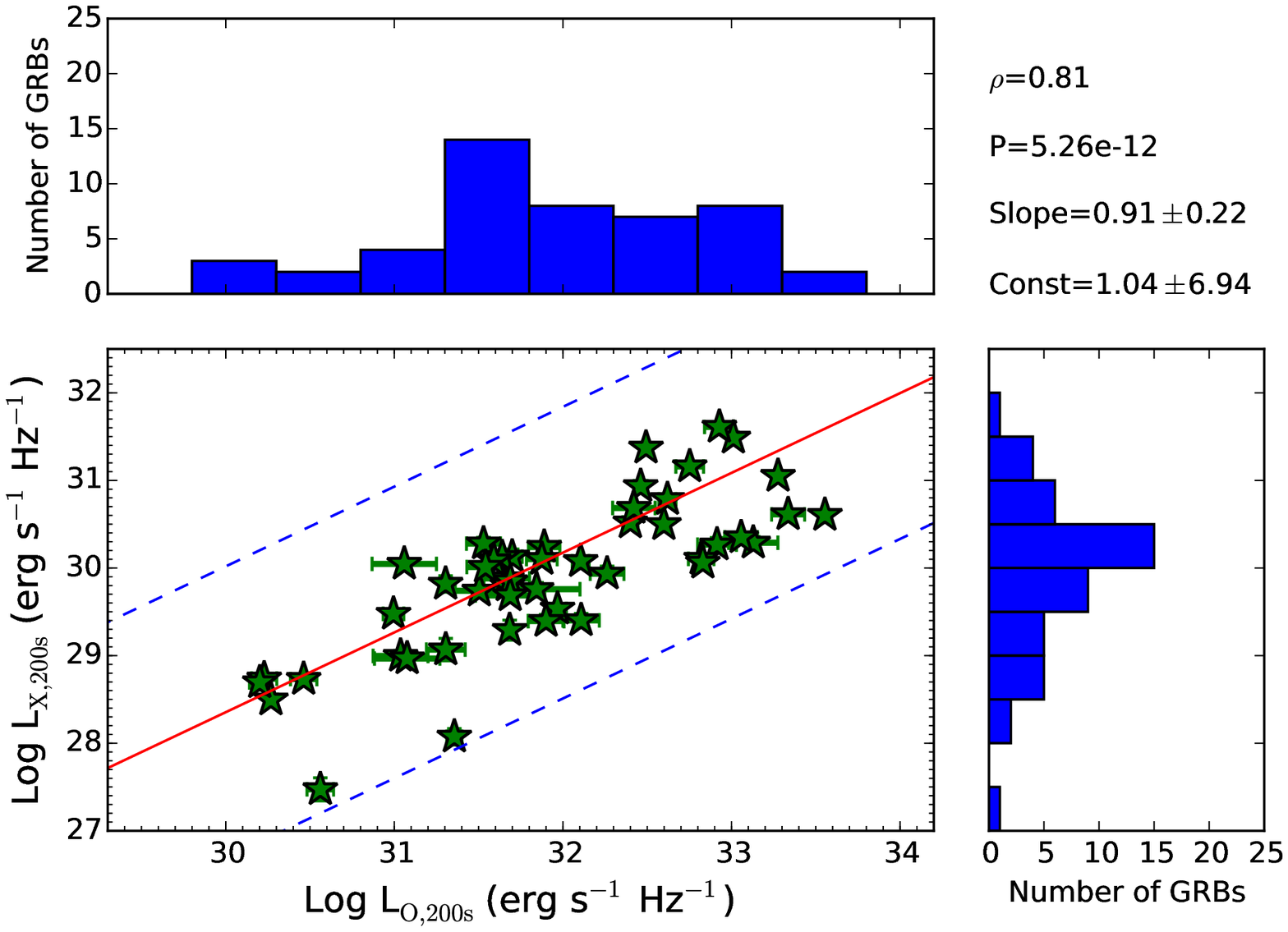}
\includegraphics[trim=0cm 0cm 0cm 5.5cm,clip=true,angle=0,scale=0.45]{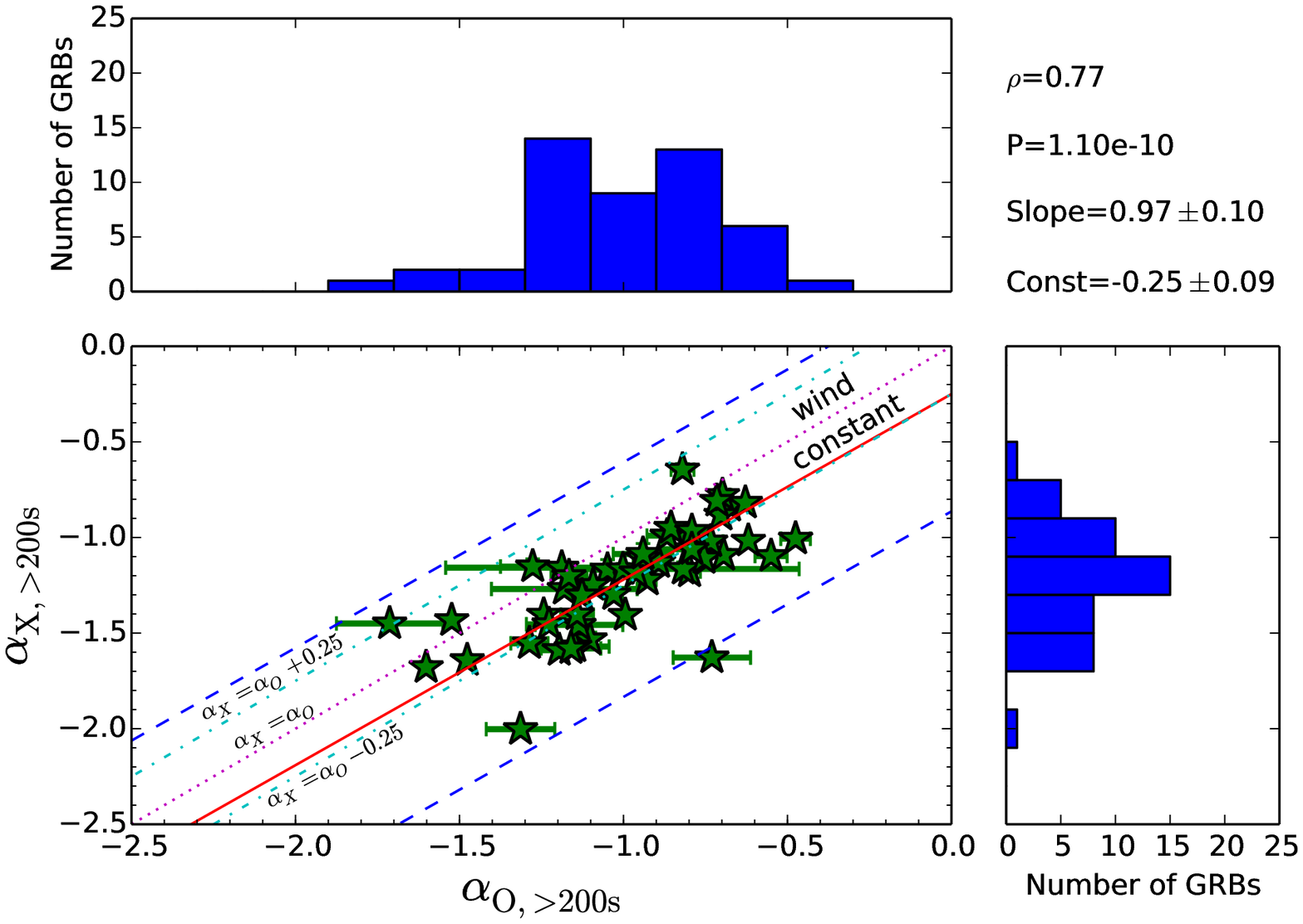}
\caption{Top: Optical/UV and X-ray luminosity determined at restframe 200~s. Bottom: 
Average decay rate of the optical/UV and X-ray light curves determined from restframe 200~s onwards. 
In both panels, the red solid line represents the best fit regression and the blue dashed line 
represents 3 times the root mean square (RMS) deviation. In the bottom panel, we also 
show relationships expected between the optical/UV and X-ray light curves from the GRB closure relations. 
The pink dotted line represents the optical/UV and X-ray decay indices being equal. The light blue dotted-dashed 
lines represent the X-ray temporal index equal to the optical/UV temporal index $\pm0.25$. In the top right corner of each panel, 
we give the Spearman rank coefficient, $\rho$, and corresponding null hypothesis probability, $P$, and we 
provide the best fit slope and constant determined by linear regression. }
\label{OX_comp}
\end{figure}

\begin{table*}
\centering
\begin{tabular}{llD{.}{.}{1.7}D{,}{\times}{1.2}D{.}{.}{1.2}D{,}{\times}{1.2}D{,}{\pm}{5.2}D{,}{\pm}{5.2}}
\hline
\multicolumn{2}{c}{Parameters} &  \multicolumn{1}{c}{Spearman Rank} & \multicolumn{1}{c}{Null} &  \multicolumn{1}{c}{Partial} &  \multicolumn{1}{c}{Null}  & \multicolumn{2}{c}{---Best fit linear regression---}\\
\multicolumn{1}{c}{$x$-axis} & \multicolumn{1}{c}{$y$-axis} & \multicolumn{1}{c}{Coefficient} & \multicolumn{1}{c}{Hypothesis} & \multicolumn{1}{c}{Spearman Rank} & \multicolumn{1}{c}{Hypothesis} & \multicolumn{1}{c}{Slope} & \multicolumn{1}{c}{Constant}\\
%[-2.0ex]
\hline
\hline
$\rm log\;L_{O,200\rm{s}}$ & $\rm log\;L_{X,200\rm{s}}$       & 0.81\;(0.05) & 5.26,10^{-12} & 0.70 & 2.85,10^{-8}  & 0.91,0.22 & 1.04,6.94\\
$\alpha_{O,>200\rm{s}}$ & $\alpha_{X,>200\rm{s}}$             & 0.77\;(0.07) & 1.10,10^{-10} & 0.75 & 1.27,10^{-9} & 0.97,0.10 & -0.25,0.09 \\[0.3cm]

$\rm log\;L_{O,200\rm{s}}$ & $\alpha_{O,>200\rm{s}}$          & -0.58\;(0.11) & 1.90,10^{-5}  & -0.50 & 2.85,10^{-4} & -0.28,0.04 & 7.72,1.31\\
$\rm log\;L_{X,200\rm{s}}$ & $\alpha_{X,>200\rm{s}}$          & -0.69\;(0.09) & 8.03,10^{-8}  & -0.63 & 1.58,10^{-6} & -0.26,0.05 & 6.71,1.39 \\

$\rm log\;L_{O,200\rm{s}}$ & $\alpha_{X,>200\rm{s}}$          & -0.60\;(0.12) & 6.87,10^{-6}  & -0.52 & 1.53,10^{-4} & -0.29,0.03 & 8.13,1.08\\
$\rm log\;L_{X,200\rm{s}}$ & $\alpha_{O,>200\rm{s}}$          & -0.65\;(0.10) & 5.58,10^{-7}  & -0.60 & 7.58,10^{-6} & -0.32,0.06 & 8.70,1.68\\[0.3cm]

$\rm {log\;E_{iso}}$ & $\alpha_{O,>200\rm{s}}$    & -0.54\;(0.12) & 9.05,10^{-5}  & -0.44 & 1.96,10^{-3} & -0.21,0.05 &  10.22,2.57 \\
$\rm {log\;E_{iso}}$ & $\alpha_{X,>200\rm{s}}$    & -0.57\;(0.11) & 3.12,10^{-5}  & -0.47 & 8.70,10^{-4} & -0.21,0.04 &  9.60,2.16 \\
$\rm {log\;E_{iso}}$ & $\rm log\;L_{O,200\rm{s}}$ &  0.76\;(0.06) & 4.51,10^{-10} &  0.66 & 4.59,10^{-7} &  1.09,0.13 & -25.27,6.92  \\
$\rm {log\;E_{iso}}$ & $\rm log\;L_{X,200\rm{s}}$ &  0.83\;(0.05) & 5.04,10^{-13} &  0.76 & 4.78,10^{-10} & 1.10,0.15 & -27.81,7.89  \\
$\rm {log\;E_{iso}}$ & $\rm log\;(L_{O,200\rm{s}}/L_{X,200\rm{s}})$ & -0.06\;(0.16) &  7.10,10^{-1} & -0.14 & 3.56,10^{-1} & -0.10,0.19 & 7.54,9.82\\[0.3cm]

$\rm {log\;E_{peak}}$  & $\alpha_{O,>200\rm{s}}$    &  -0.45\;(0.13) & 2.05,10^{-3} & -0.38 & 1.20,10^{-2} &-0.48,0.17 &  0.22,0.41 \\
$\rm {log\;E_{peak}}$  & $\alpha_{X,>200\rm{s}}$    &  -0.48\;(0.13) & 9.22,10^{-4} & -0.40 & 7.52,10^{-3} &-0.48,0.15 & -0.03,0.36 \\
$\rm {log\;E_{peak}}$  & $\rm log\;L_{O,200\rm{s}}$ &   0.66\;(0.11) & 1.16,10^{-6} &  0.58 & 3.51,10^{-5} & 2.97,0.76 & 24.53,1.95 \\
$\rm {log\;E_{peak}}$  & $\rm log\;L_{X,200\rm{s}}$ &   0.75\;(0.10) & 4.74,10^{-9} &  0.70 & 1.38,10^{-7} & 2.97,0.67 & 22.50,1.73 \\[0.3cm]

$\rm {log\;T90_{rest}}$   & $\alpha_{O,>200\rm{s}}$    & -0.23\;(0.14) & 1.15,10^{-1}  & -0.21 & 1.61,10^{-1} & -0.19,0.10 & -0.75,0.12\\
$\rm {log\;T90_{rest}}$   & $\alpha_{X,>200\rm{s}}$    & -0.13\;(0.14) & 3.71,10^{-1}  & -0.10 & 5.03,10^{-1} & -0.12,0.09 & -1.08,0.11\\
$\rm {log\;T90_{rest}}$   & $\rm log\;L_{O,200\rm{s}}$ &  0.26\;(0.14) & 7.58,10^{-2}  &  0.24 & 9.85,10^{-2} &  4.41,2.59 & 26.28,3.49\\
$\rm {log\;T90_{rest}}$   & $\rm log\;L_{X,200\rm{s}}$ &  0.14\;(0.15) & 3.58,10^{-1}  &  0.09 & 5.58,10^{-1} &  10.60,12.99 & 16.35,16.70\\
$\rm {log\;T90_{rest}}$   & $\rm {log\;E_{iso}}$      &  0.43\;(0.12) & 2.65,10^{-3}  &  0.43 & 2.34,10^{-3} &  2.30,0.57 & 49.70,0.80\\
$\rm {log\;T90_{rest}}$   & $\rm {log\;E_{peak}}$      &  0.23\;(0.16) & 1.26,10^{-1}  &  0.21 & 1.69,10^{-1} &  0.46,0.21 & 1.90,0.25 \\
\hline
\end{tabular}
\caption{For each pair of parameters examined, this table provides: the Spearman rank correlation coefficient with its associated null hypothesis; 
the coefficient of the partial Spearman rank with its associated null hypothesis, which tests the correlation between two parameters taking into 
account the parameters dependence on redshift; the slope and constant values provided by the best fit linear regression. For comparison with our 
Monte Carlo simulations in \S\ref{discussion}, we also provide the $1\sigma$ error of the Spearman rank coefficient.}
\label{Spear_lin}
\end{table*}

\section{Observational Results}
\label{results}
In this section we now determine what correlations we find in the observed sample of 48 GRB afterglows. We use
the Spearman rank correlation to determine if two parameters are correlated and linear regression to quantify the degree of correlation
and the relationship between these parameters. The results of the Spearman rank tests and linear 
regressions for all correlations can be found in Table \ref{Spear_lin}. In this table we also include the partial Spearman rank 
correlation, which measures the degree of correlation between two parameters, excluding the effect of a third, in this case redshift 
\citep[see][for further details]{ken79}.

In \cite{oates12}, a correlation was discovered between the logarithmic luminosity at 200~s and the average decay rate of optical/UV light
curves, measured from 200~s onwards. We now examine if a similar relationship is observed in the X-ray light curves. The best fits to the
linear regressions for these correlations: $\rm log\;L_{X,200\rm{s}}-\alpha_{X,>200\rm{s}}$ and $\rm log\;L_{O,200\rm{s}}-\alpha_{O,>200\rm{s}}$
can be found in Table \ref{Spear_lin}. Similar to that found in the optical/UV, we find a significant relationship between the luminosity
and decay rate of the X-ray light curves \citep[see also][]{racusin15}. For the two frequencies, we find the linear regression gives
relationships that are consistent at $1\sigma$.

The X-ray light curves in our sample display a wide range in behaviour, ranging from X-ray light curves
with simple power-law decay to GRBs with 4-5 breaks. Since \cite{dai08} have shown a correlation between the luminosity at the
end of the X-ray plateau with the time the plateau phase ceases, we examined whether the plateau phase plays a role in our correlation.
In our sample we find 35 GRBs that have a plateau following the criteria of \cite{racusin09}. Repeating the Spearman rank correlation
for $\rm log\;L_{X,200\rm{s}}-\alpha_{X,>200\rm{s}}$ for these GRBs, we find a coefficient of -0.73 at a confidence of $4.97\sigma$,
indicating a strong correlation. For the 13 GRBs without X-ray plateau phase, repeating the Spearman rank correlation, we find a
coefficient of -0.32 at 72 per cent confidence. It is inconclusive whether or not there is a correlation between $\rm log\;L_{X,200\rm{s}}-\alpha_{X,>200\rm{s}}$
for the GRBs without X-ray plateau phase. However considering for the full sample of GRBs, the X-ray and optical/UV light curves have
consistent linear regressions (the optical light curves do not typically have a well defined plateau phase), suggests that the
correlation does not depend on having a plateau phase in the optical or X-ray light curves. This is also supported by the results
of a similar comparison using a larger X-ray sample \citep{racusin15}. Therefore in this paper we do not distinguish further
between GRBs with and without X-ray plateaus.

Since Eq. \ref{lum_eq} predicts a relationship between luminosity of the X-ray and optical/UV afterglows and we find correlations 
that show that intrinsically the brightest afterglows decay the quickest, we now examine and compare the 
$\rm log\;L_{200\rm{s}}$ and $\alpha_{>200\rm{s}}$ parameters of the optical/UV and X-ray light curves.

\subsection{Afterglow Parameter Comparison}
In the top panel of Fig. \ref{OX_comp}, we compare the optical/UV and X-ray luminosities at 200~s,  
$\rm log\;L_{O,200\rm{s}}$ and $\rm log\;L_{X,200\rm{s}}$, respectively. There is a strong positive correlation 
between the luminosity in the X-ray and optical/UV bands at 200~s, which is confirmed by a Spearman rank correlation 
coefficient of 0.81 at a significance of 6.9$\sigma$. A linear regression of the two parameters results in a relationship 
close to unity, with the slope $0.91\pm0.22$.

\begin{figure*}
\includegraphics[trim=0cm 0cm 0cm 5.5cm,clip=true,angle=0,scale=0.35]{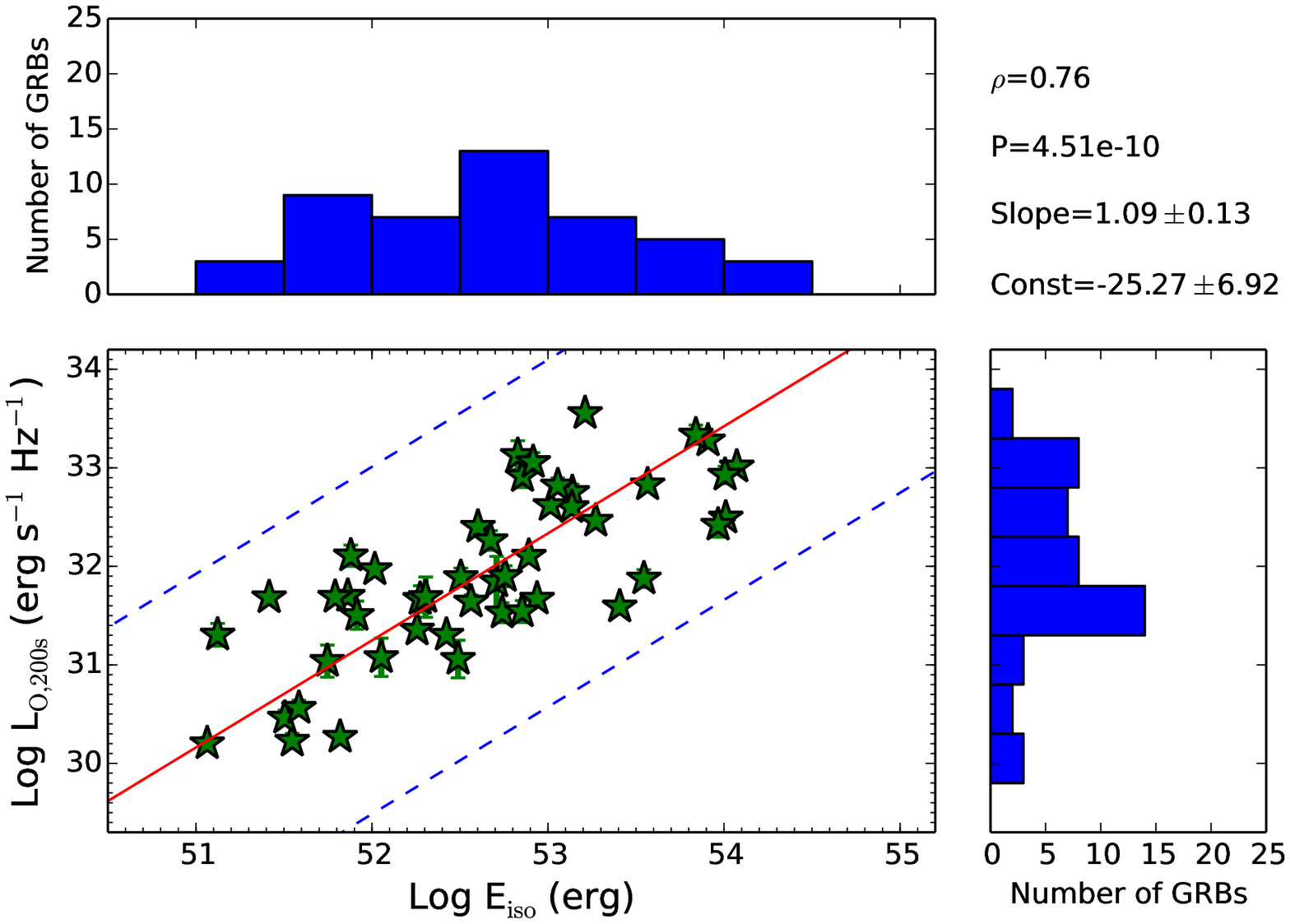}
\includegraphics[trim=0cm 0cm 0cm 5.5cm,clip=true,angle=0,scale=0.35]{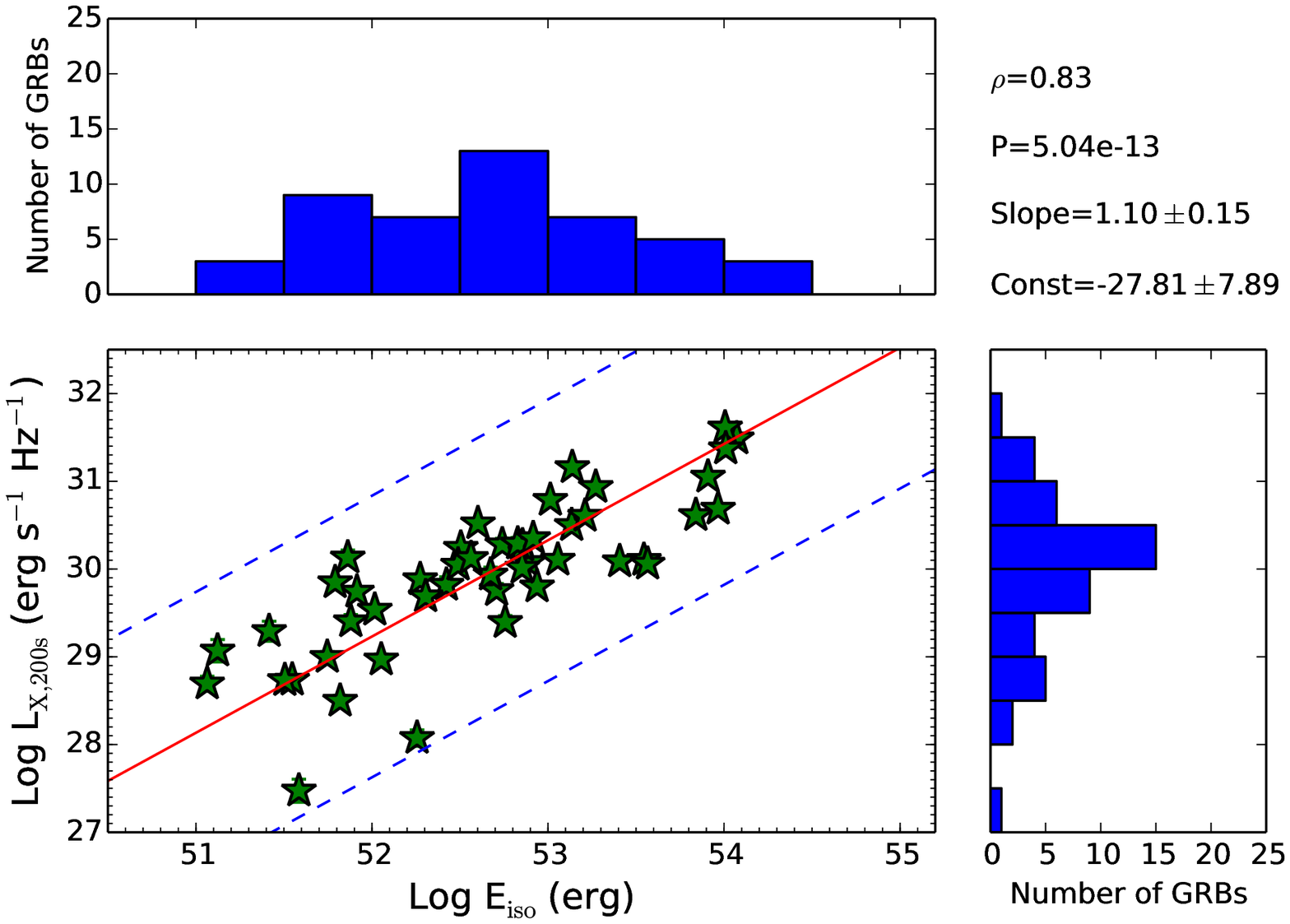}
\includegraphics[trim=0cm 0cm 0cm 5.5cm,clip=true,angle=0,scale=0.35]{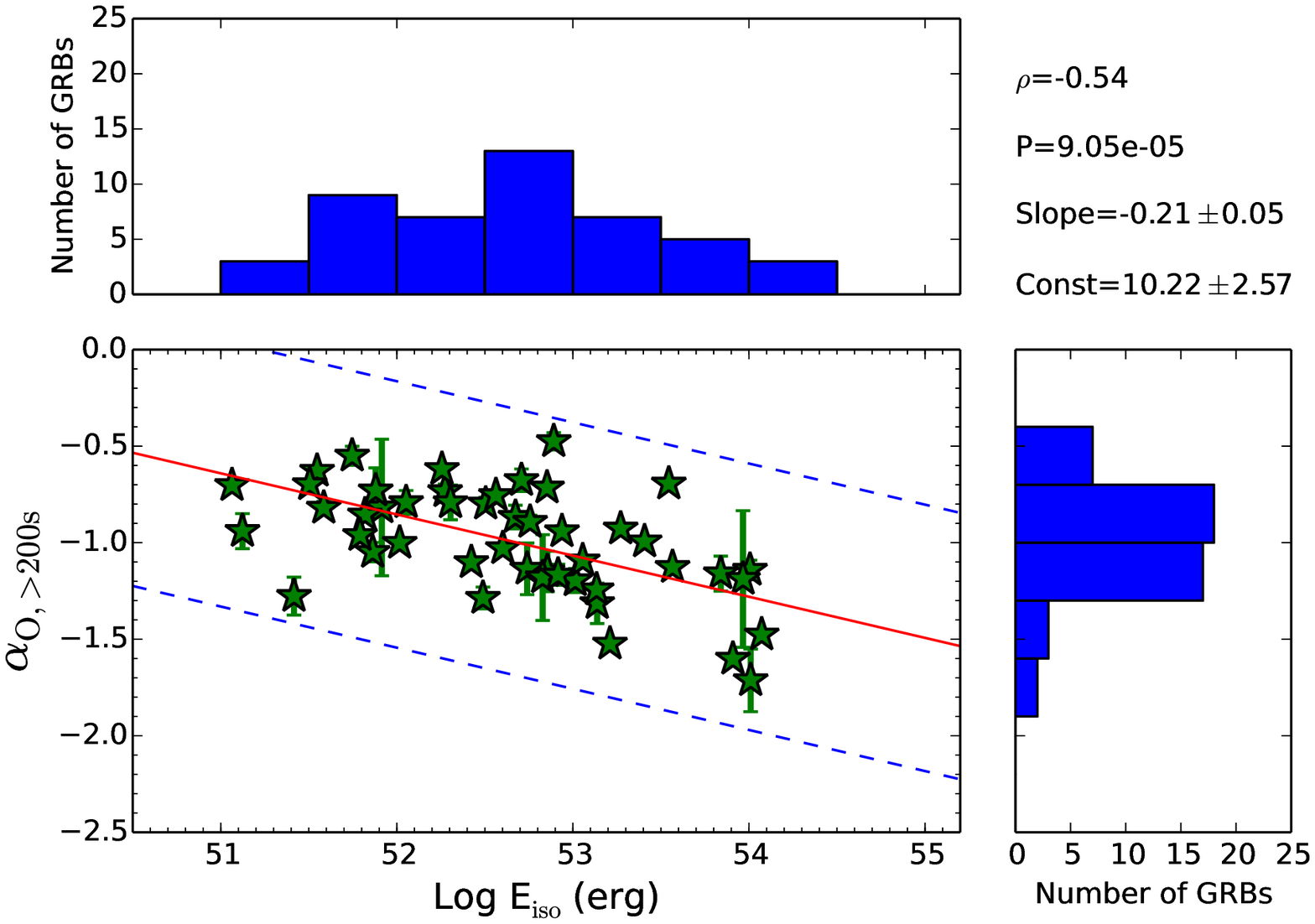}
\includegraphics[trim=0cm 0cm 0cm 5.5cm,clip=true,angle=0,scale=0.35]{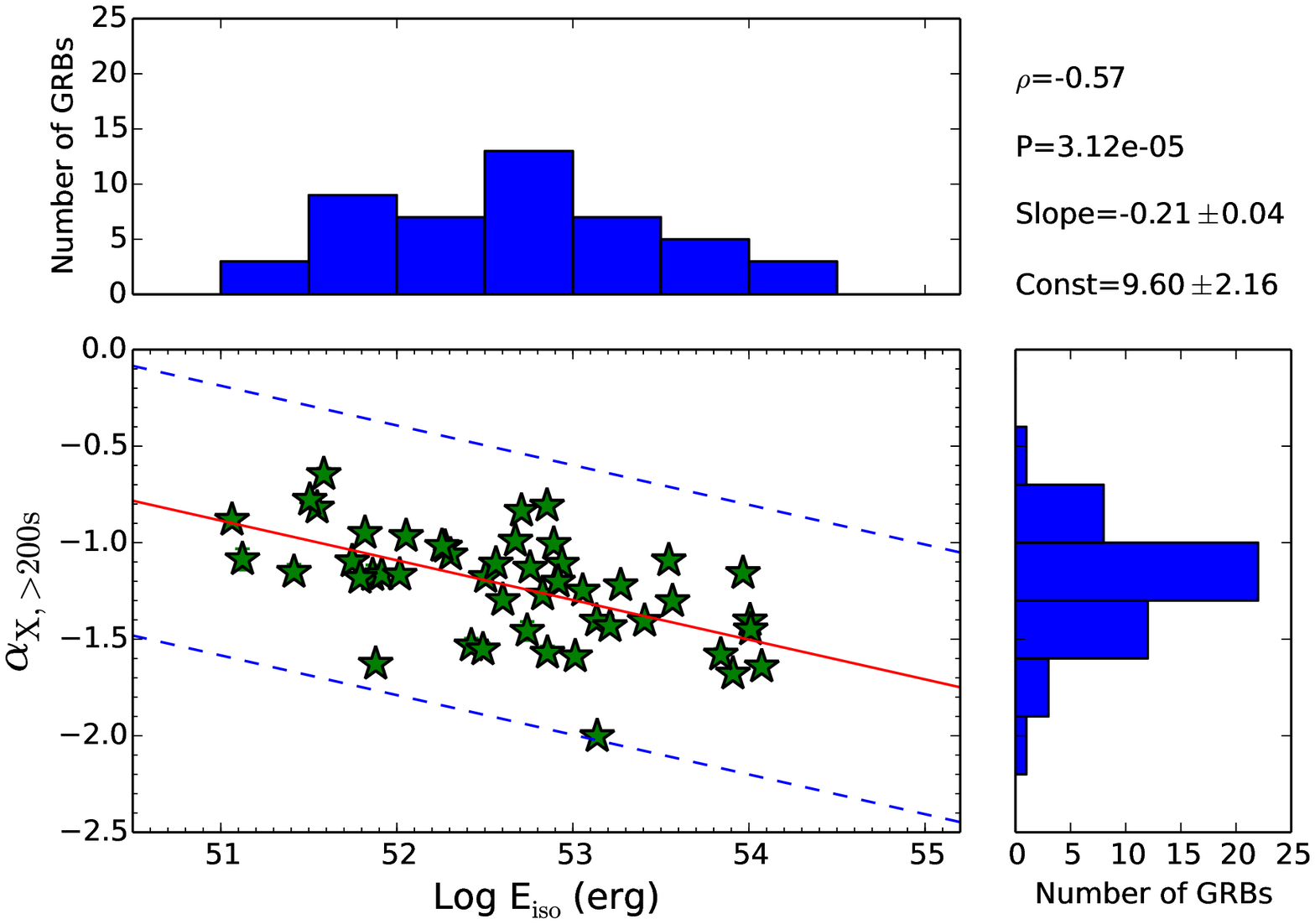}
\caption{Top Left: The optical luminosity at restframe 200~s versus isotropic energy. 
Top Right: The X-ray luminosity at restframe 200~s versus isotropic energy.
Bottom Left: The optical average decay index determined from restframe 200~s versus isotropic energy. 
Bottom Right: The X-ray average decay index determined from restframe 200~s versus isotropic energy. 
In all panels, the red solid line represents the best fit regression and the blue dashed line 
represents the 3$\sigma$ deviation. In the top right corner of each panel, 
we give the Spearman rank coefficient, $\rho$, and corresponding null hypothesis probability, P, and we 
provide the best fit slope and constant determined by linear regression.}
\label{prompt_eiso}
\end{figure*}

In the bottom panel of Fig. \ref{OX_comp}, we compare the average decay rate determined from 200~s 
onwards of the optical/UV and X-ray light curves, $\alpha_{O,>200\rm{s}}$ and $\alpha_{X,>200\rm{s}}$, respectively. Spearman 
rank gives a correlation coefficient of 0.77 at a significance of 6.5$\sigma$. 

In Table \ref{Spear_lin}, we also provide the relationships derived when swapping the X-ray and optical 
luminosity decay parameters, i.e $\rm log\;L_{O,200\rm{s}}$ versus $\alpha_{X,>200\rm{s}}$ and 
$\rm log\;L_{X,200\rm{s}}$ versus $\alpha_{O,>200\rm{s}}$. The fact that significant correlations are 
found even when mixing decay and luminosity parameters between the optical/UV and X-ray bands provides 
support to the correlations discussed in this section.

\subsection{Prompt emission and afterglow parameter comparison}
In the following we examine the relationship between $\rm log\;L_{200\rm{s}}$ and $\alpha_{>200\rm{s}}$
with $\rm log\;E_{iso}$ for both the optical/UV and X-ray light curves, so that we may compare the observed
correlations with our simulations. We will also compare the afterglow parameters with other basic properties of the prompt emission. 

\subsubsection{Prompt emission: isotropic energy}
Comparisons between afterglow luminosity and isotropic energy have been previously reported. Early reports showed
 that the luminosity at ~12~hours to 1~day after the trigger correlates well with the isotropic $\gamma$-ray energy 
\citep[e.g.][]{kou04,depas06,nys09,kan10}. More recently \cite{dav12} and \cite{mar13}, showed that in the X-rays, 
measurements of the luminosity during the early afterglow, approximately $5-10$ minutes after trigger, have less scatter 
and the correlations are stronger in comparison with measurements taken at any subsequent time later. We may also thus 
expect this to be the case in the optical. In the top two panels of Fig. \ref{prompt_eiso}, we display the logarithmic 
isotropic $\gamma$-ray energy, $\rm log\;E_{iso}$ against $\rm log\;L_{O,200\rm{s}}$ and $\rm log\;L_{X,200\rm{s}}$. 
Spearman rank correlations of the luminosity parameters against $\rm log\;E_{iso}$ provide correlation coefficients 
of 0.76 and 0.83 with significances of $6.2\sigma$ and $7.2\sigma$ for the optical/UV and X-ray afterglows, respectively. 
In comparison to \cite{nys09} and \cite{kan10}, who compare the optical/UV luminosity with $\rm E_{iso}$ at 11~hours and 1~day, 
respectively, we see less spread using the luminosity at earlier times, as expected in comparison to the X-ray light curves.
For both the optical/UV and X-ray light curves, the linear regressions of the $\rm log\;L_{200\rm{s}}$ and $\rm log\;E_{iso}$, give 
consistent results within $1\sigma$ errors. 

\begin{figure}
\includegraphics[trim=0cm 0cm 0cm 5.5cm,clip=true,angle=0,scale=0.45]{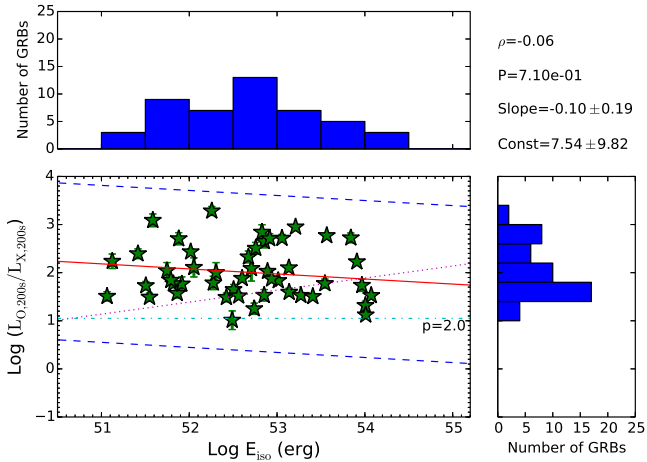}
\caption{The ratio of the optical/UV to X-ray luminosities at 200~s versus the isotropic prompt emission $\rm E_{iso}$. The 
red solid line represents the best fit regression and the blue dashed line 
represents the 3$\sigma$ deviation. The dotted pink line represents how the ratio scales with energy in the regime 
$\nu_m<\nu_O<\nu_c<\nu_X$ as given in the second line of Eq. \ref{rat_energy_eq}; the line has arbitrary normalization. 
The dotted-dashed light blue line, indicates the minimum ratio value predicted from Eq. \ref{rat_energy_eq} when $p$ is 2.0.
In the top right corner, we give the Spearman rank coefficient, $\rho$, and corresponding null hypothesis probability, P, and we 
provide the best fit slope and constant determined by linear regression.}
\label{OX_rat}
\end{figure}

In Fig. \ref{OX_rat} we display the relationship between $\rm log\;L_{O,200\rm{s}}$ and $\rm log\;L_{X,200\rm{s}}$
with $\rm log\;E_{iso}$, as the logarithm of the ratio between the optical/UV and X-ray luminosities against $\rm log\;E_{iso}$
(see Eq. \ref{rat_energy_eq}). There is no evidence for correlation between these parameters.

In the bottom two panels of Fig. \ref{prompt_eiso} we display $\rm log\;E_{iso}$ against $\alpha_{O,>200\rm{s}}$ 
and $\alpha_{X,>200\rm{s}}$. Both panels indicate correlations between the average decay indices with 
isotropic energy. This suggests that the more energetic the prompt emission, the faster the average decay of the X-ray and 
optical/UV afterglows. Spearman rank correlations of the average decay parameters against $\rm log\;E_{iso}$ provide correlation 
coefficients of -0.54 and -0.57 for the optical/UV and X-ray afterglows, at confidences of $3.9\sigma$ and $4.2\sigma$ respectively. 
These correlations are slightly less strong in comparison to that found between the luminosity and $\rm log\;E_{iso}$. We 
note that within errors the equations for the linear regression for both the optical/UV and X-ray decay indices 
against $\rm log\;E_{iso}$ are consistent with each other.

\subsubsection{Prompt emission: peak spectral energy}
The Amati relation indicates a relationship between the isotropic $\gamma$-ray energy $\rm E_{iso}$ and the restframe 
$\gamma$-ray peak energy $\rm E_{peak}$ \citep{ama02}. Therefore we may already predict correlations between
$\rm E_{peak}$ and the afterglow parameters, but for completeness and to report the strength of these correlations we
now briefly compare the afterglow parameters with $\rm E_{peak}$.

\begin{figure*}
\includegraphics[trim=0cm 0cm 0cm 5.5cm,clip=true,angle=0,scale=0.35]{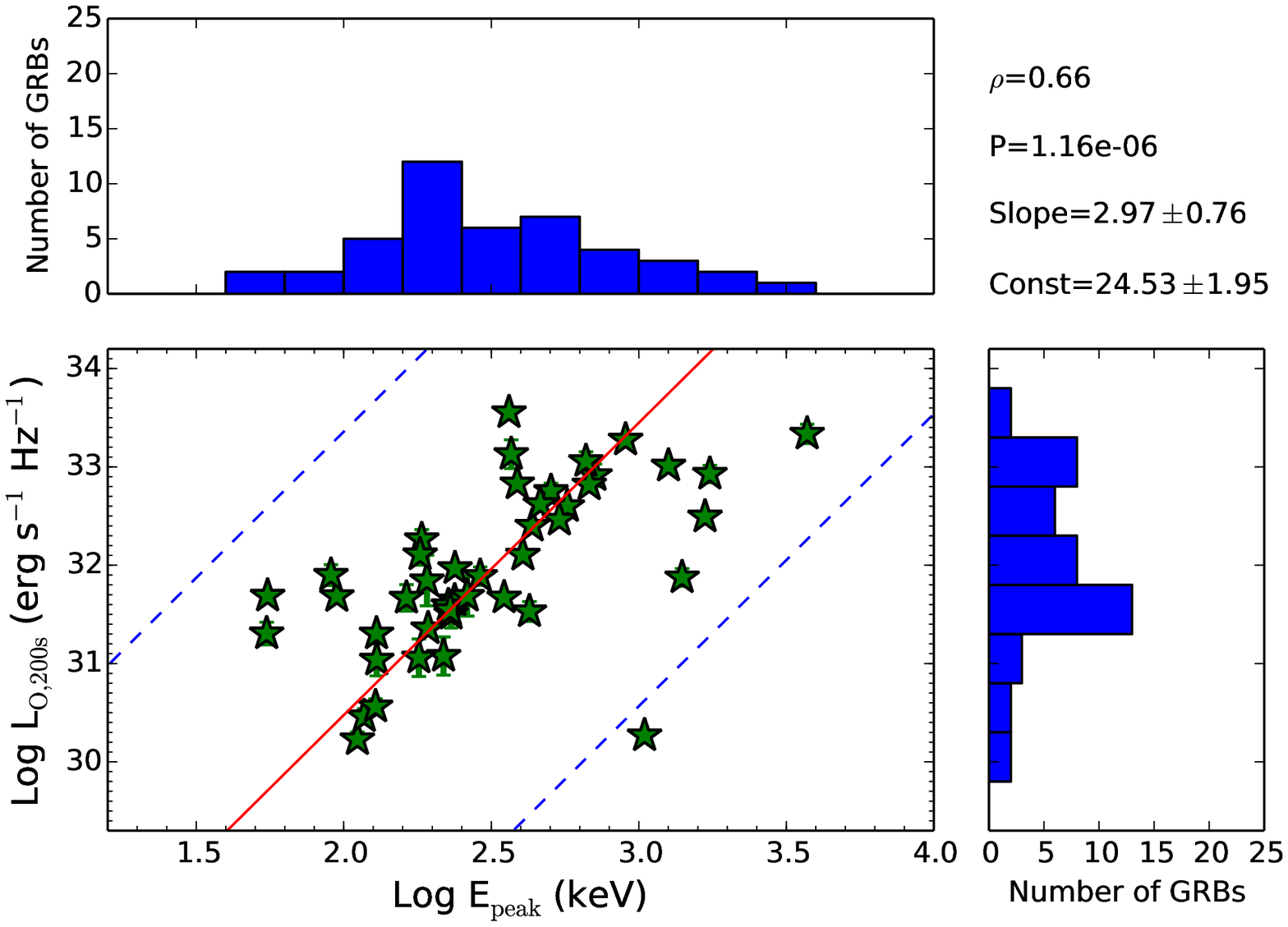}
\includegraphics[trim=0cm 0cm 0cm 5.5cm,clip=true,angle=0,scale=0.35]{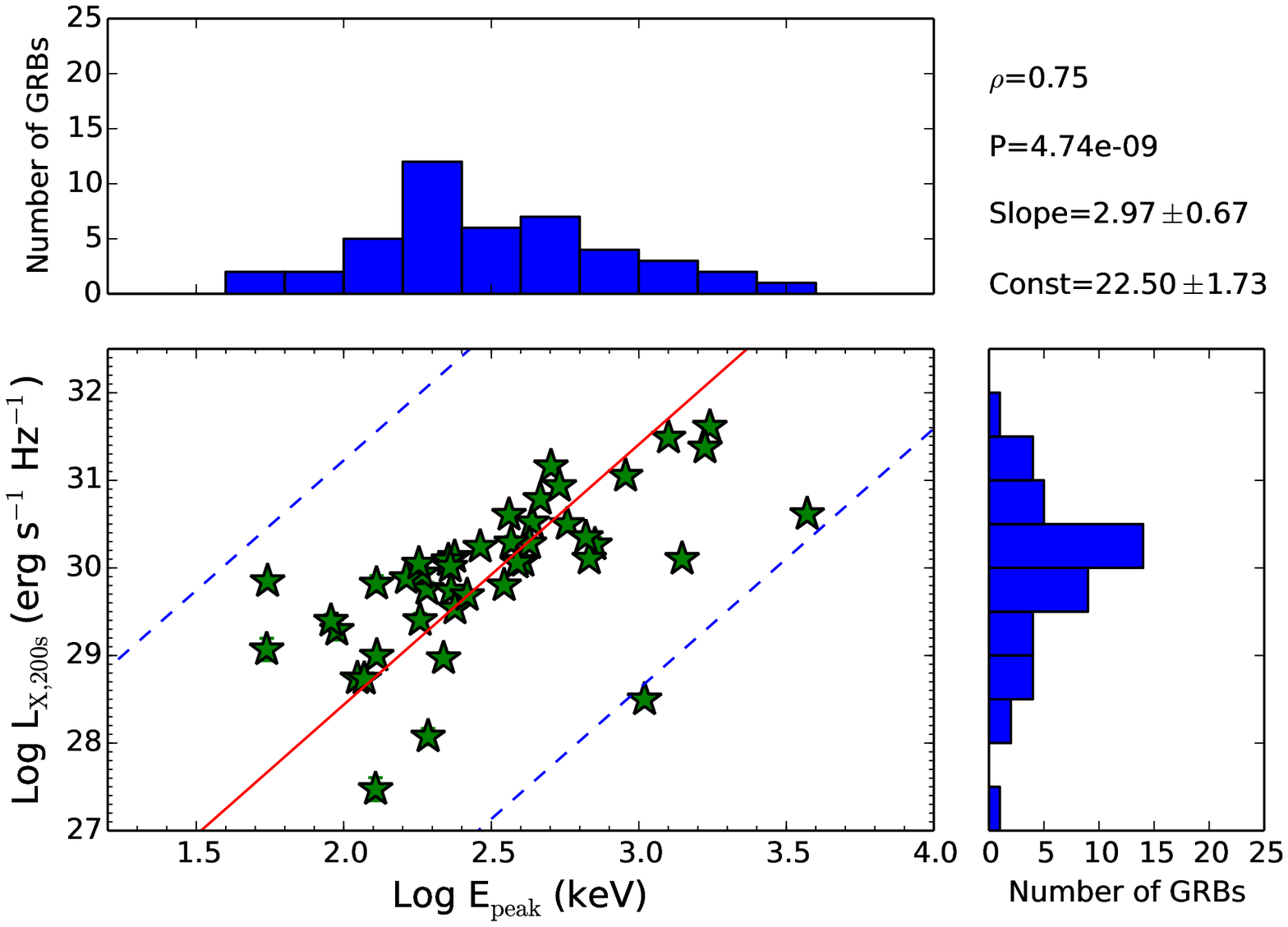}
\includegraphics[trim=0cm 0cm 0cm 5.5cm,clip=true,angle=0,scale=0.35]{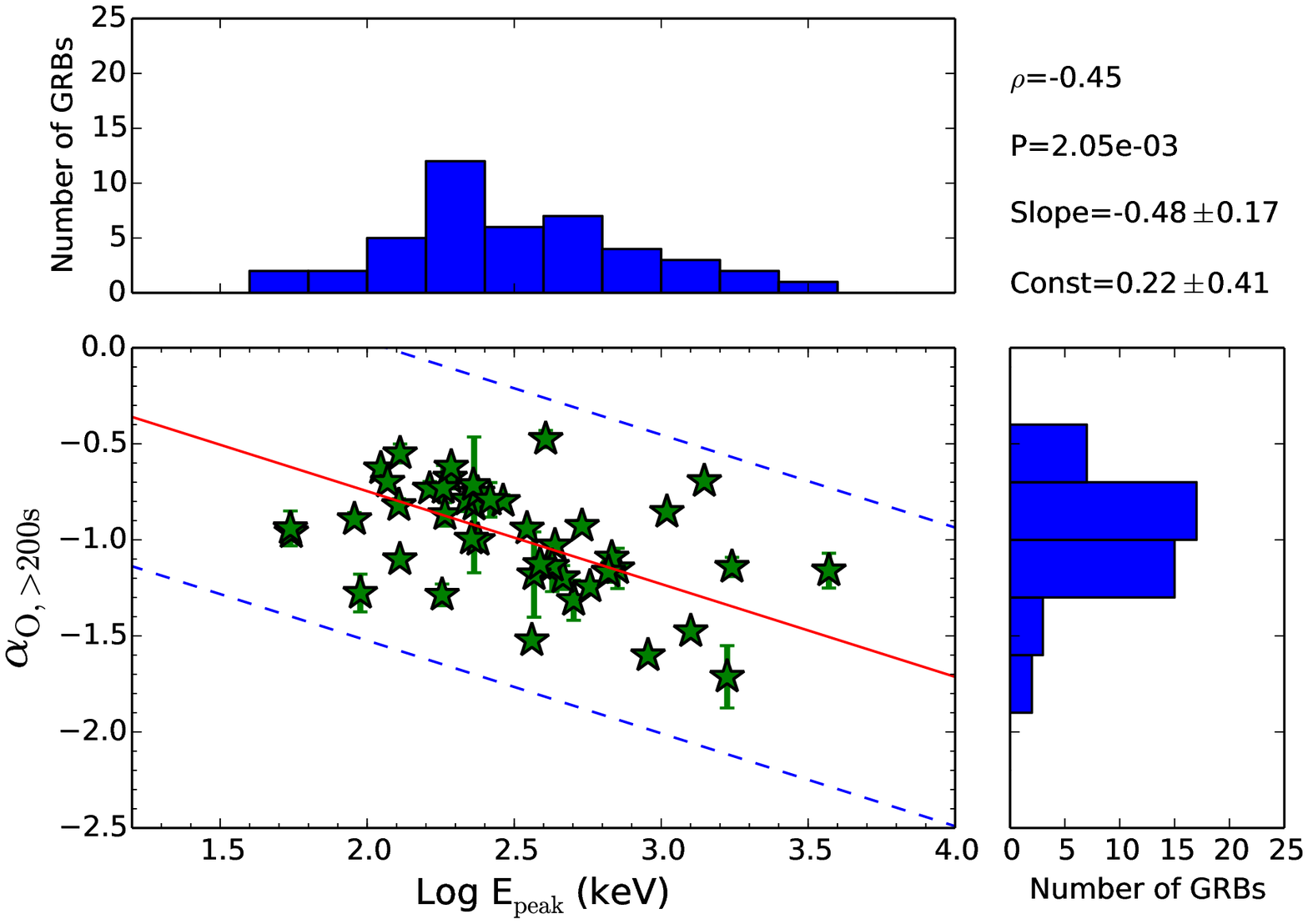}
\includegraphics[trim=0cm 0cm 0cm 5.5cm,clip=true,angle=0,scale=0.35]{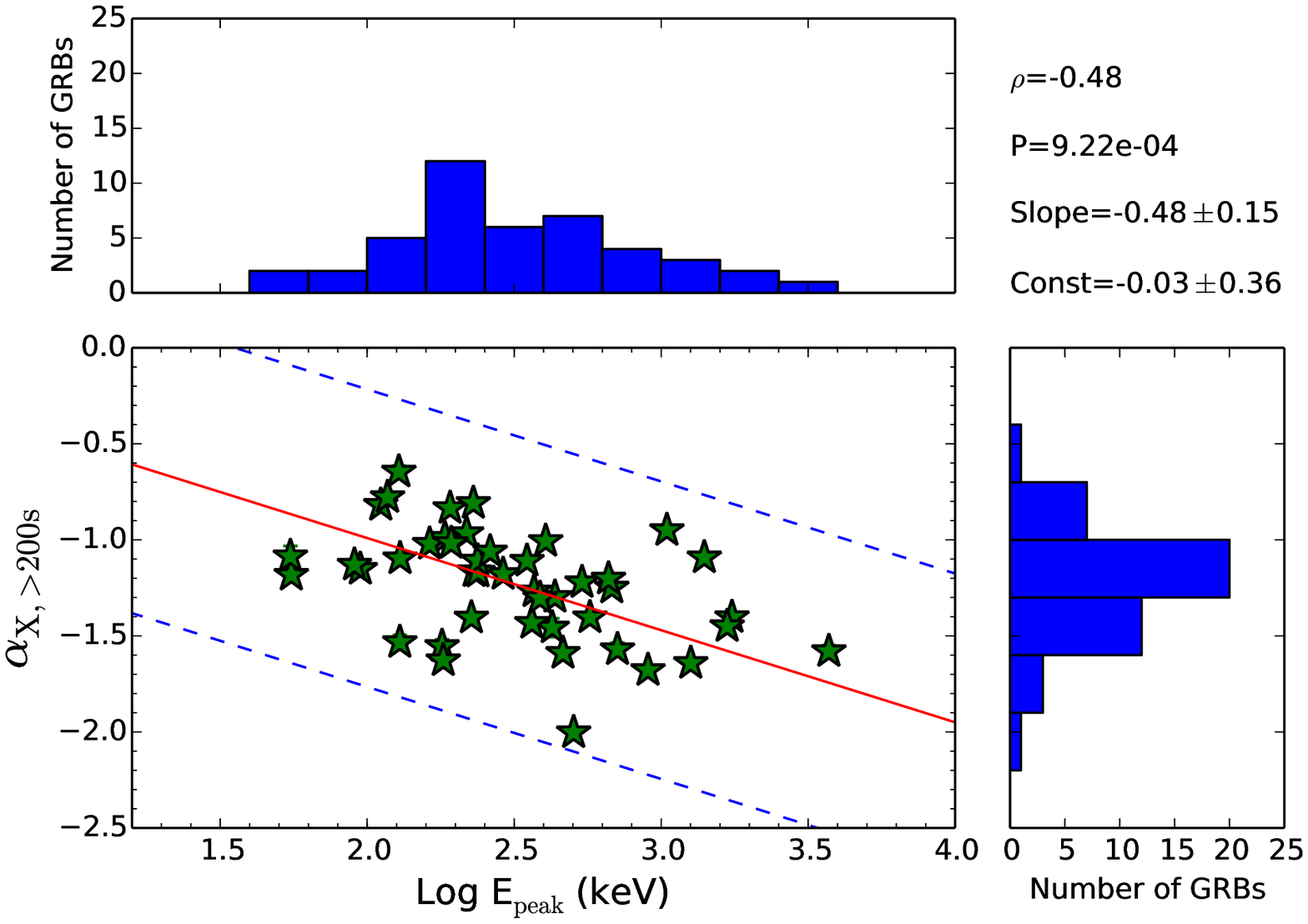}
\caption{Top Left: The optical luminosity at restframe 200~s versus $\gamma$-ray peak energy. 
Top Right: The X-ray luminosity at restframe 200~s versus $\gamma$-ray peak energy.
Bottom Left: The optical average decay index determined from restframe 200~s versus $\gamma$-ray peak energy. 
Bottom Right: The X-ray average decay index determined from restframe 200~s versus $\gamma$-ray peak energy. 
In all panels, the red solid line represents the best fit regression and the blue dashed line 
represents the 3$\sigma$ deviation. In the top right corner of each panel, 
we give the Spearman rank coefficient, $\rho$, and corresponding null hypothesis probability, P, and we 
provide the best fit slope and constant determined by linear regression. }
\label{prompt_epeak}
\end{figure*}

In the top panels of Fig. \ref{prompt_epeak}, we display the logarithmic restframe peak $\gamma$-ray energy, $\rm log\;E_{peak}$ against 
$\rm log\;L_{O,200\rm{s}}$ and $\rm log\;L_{X,200\rm{s}}$. Spearman rank correlations of the luminosity 
parameters against $\rm log\;E_{peak}$ provide evidence for correlation with coefficients of 0.75 and 0.66 for the X-ray and 
optical/UV light curves, respectively, with corresponding significances of $5.9\sigma$ and $4.9\sigma$. This is consistent with
\cite{dav12} who also show that the early X-ray luminosity (at restframe 5 minutes) and $\rm log\;E_{peak}$ are correlated.
We notice that the Spearman rank coefficient for the $\rm log\;L_{200\rm{s}}$ versus $\rm log\;E_{peak}$ is smaller than that
found for $\rm log\;L_{200\rm{s}}$ with $\rm log\;E_{iso}$, indicating that the relationships involving the prompt emission peak energy are
weaker in comparison to the relationships observed with the isotropic energy. This is also consistent with \cite{dav12}
who find that the correlations between the X-ray luminosity and $\rm log\;E_{peak}$ have smaller correlation coefficients
than the correlation between X-ray luminosity and $\rm log\;E_{iso}$.

The bottom panels of Fig. \ref{prompt_epeak}, display $\rm log\;E_{peak}$ against $\alpha_{O,>200\rm{s}}$ and 
$\alpha_{X,>200\rm{s}}$. For these correlations, the Spearman rank coefficients are smaller 
in comparison to the Spearman rank coefficients found for the correlations between the decay indices and 
$\rm log\;E_{iso}$. The correlation of the decay indices with $\rm log\;E_{peak}$ results in coefficients of 
-0.45 and -0.48 for the optical/UV and X-ray afterglows, respectively, with corresponding significances of 
$3.1\sigma$ and $3.3\sigma$.

\subsubsection{Prompt emission: restframe T90 duration}
In Table \ref{Spear_lin} and Fig. \ref{restT90}, we also provide results of the Spearman rank correlation for the duration 
of the $\gamma$-ray emission in the restframe, ${\rm T90}/(1+z)$, with the optical/UV and X-ray 
$\rm log\;L_{200\rm{s}}$ and $\alpha_{>200\rm{s}}$ parameters. Significant correlations are not found 
amongst these parameters. This is consistent with \cite{mar13} who do not find any evidence for
correlations amongst several X-ray parameters with restframe ${\rm T90}$.

\subsection{Effect of prompt emission contaminating afterglow light curves at restframe 200s}
\label{contam}
\begin{figure*}
\includegraphics[trim=0cm 0cm 0cm 5.5cm,clip=true,angle=0,scale=0.35]{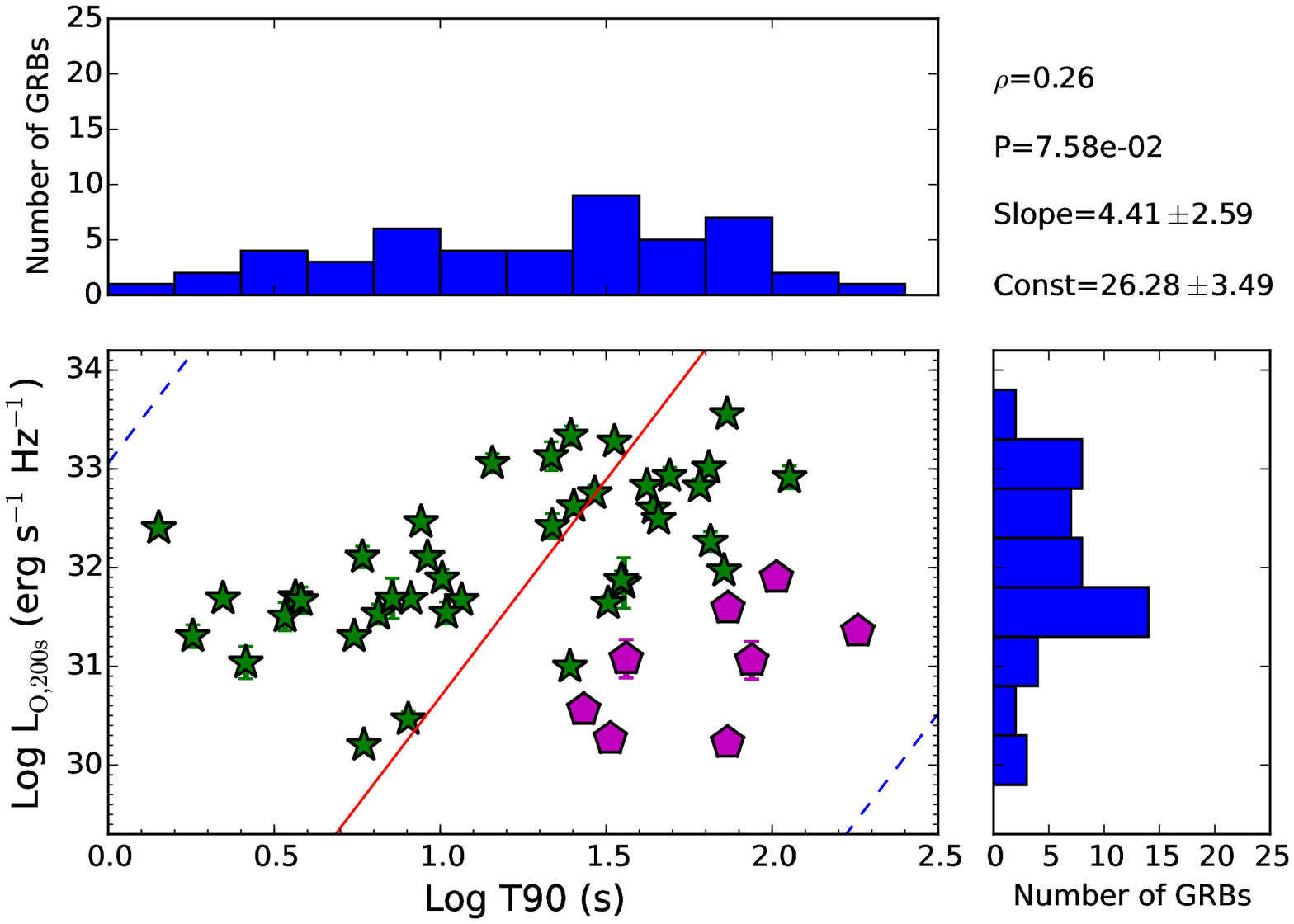}
\includegraphics[trim=0cm 0cm 0cm 5.5cm,clip=true,angle=0,scale=0.35]{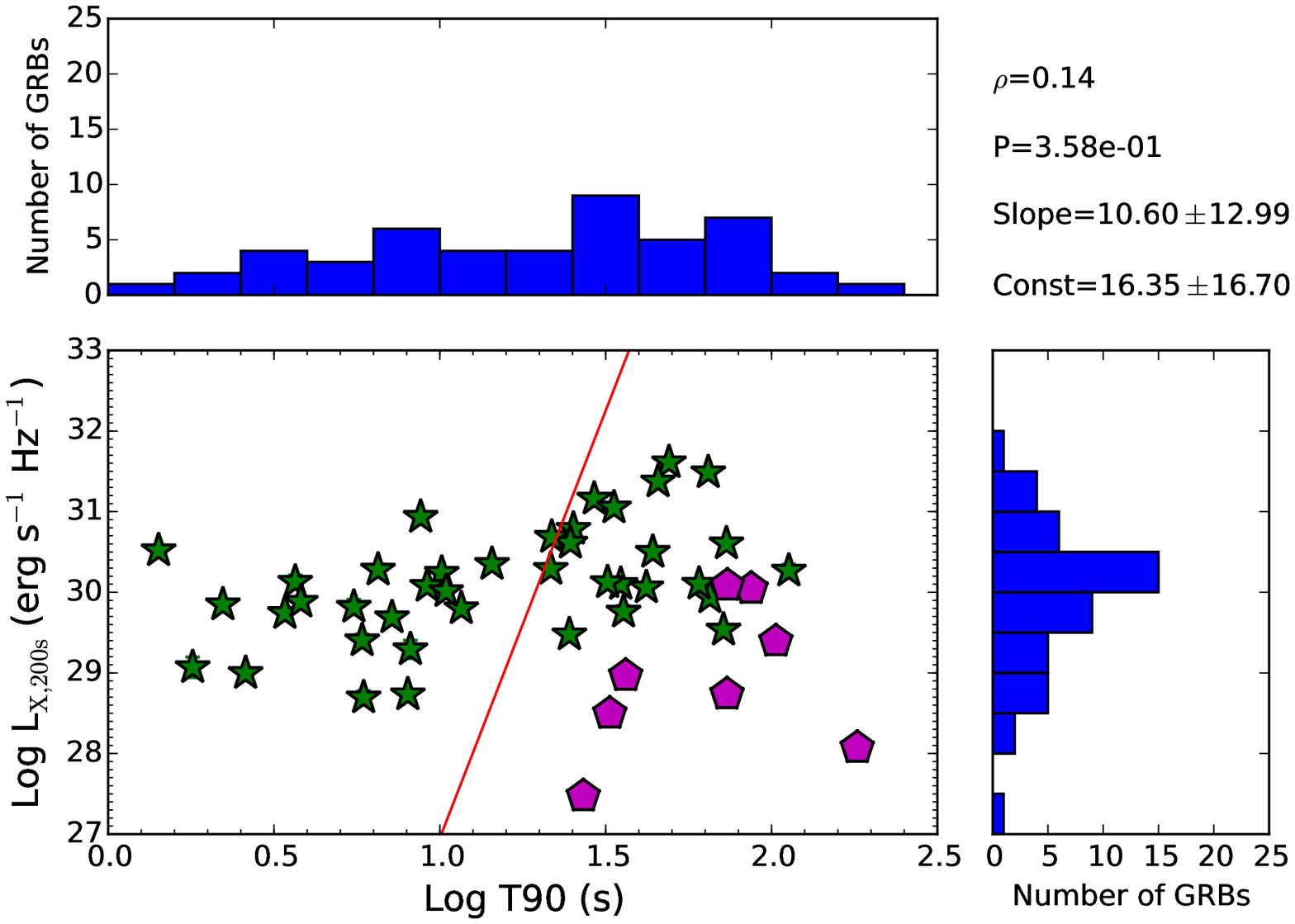}
\includegraphics[trim=0cm 0cm 0cm 5.5cm,clip=true,angle=0,scale=0.35]{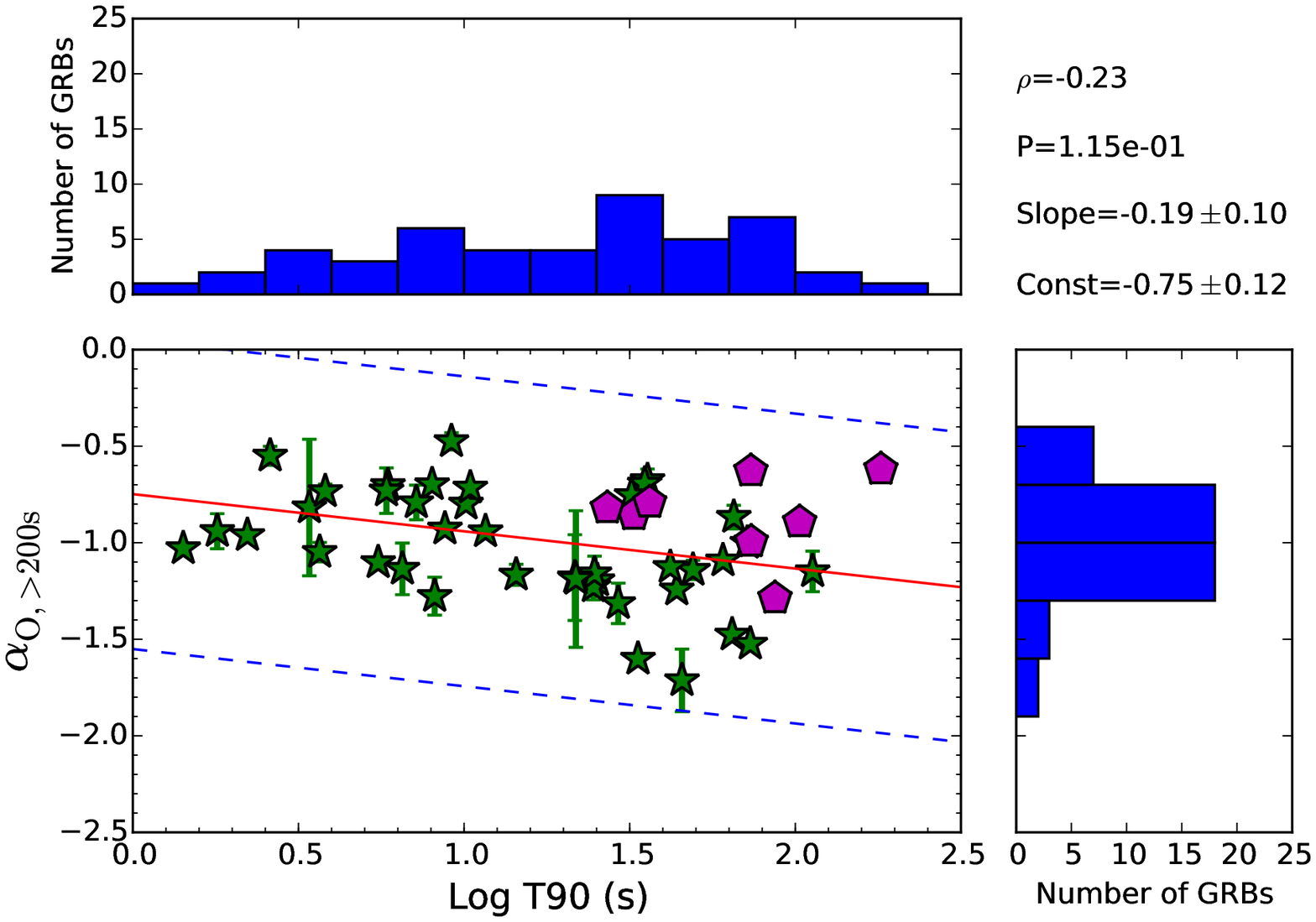}
\includegraphics[trim=0cm 0cm 0cm 5.5cm,clip=true,angle=0,scale=0.35]{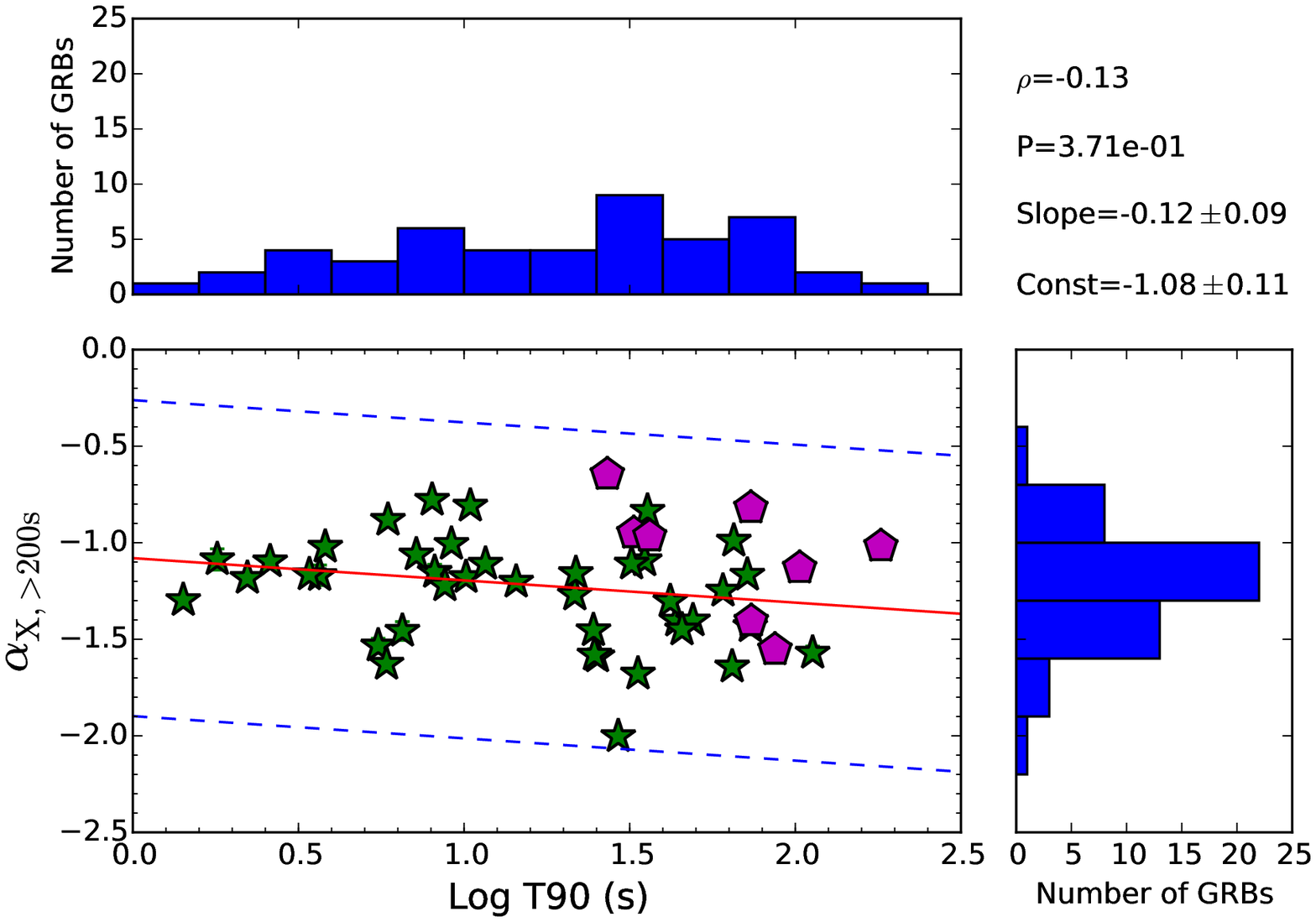}
\includegraphics[trim=0cm 0cm 0cm 5.5cm,clip=true,angle=0,scale=0.35]{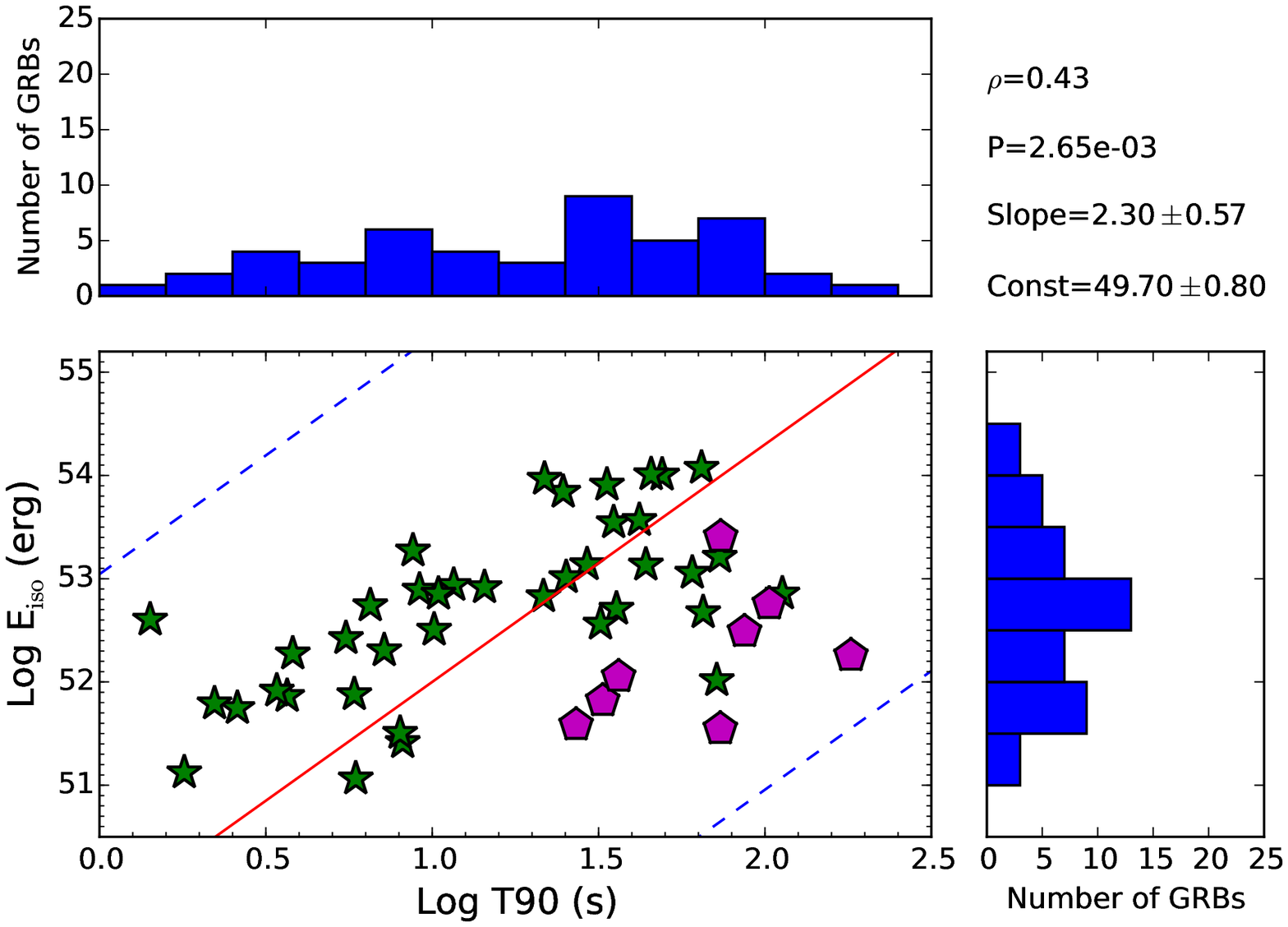}
\includegraphics[trim=0cm 0cm 0cm 5.5cm,clip=true,angle=0,scale=0.35]{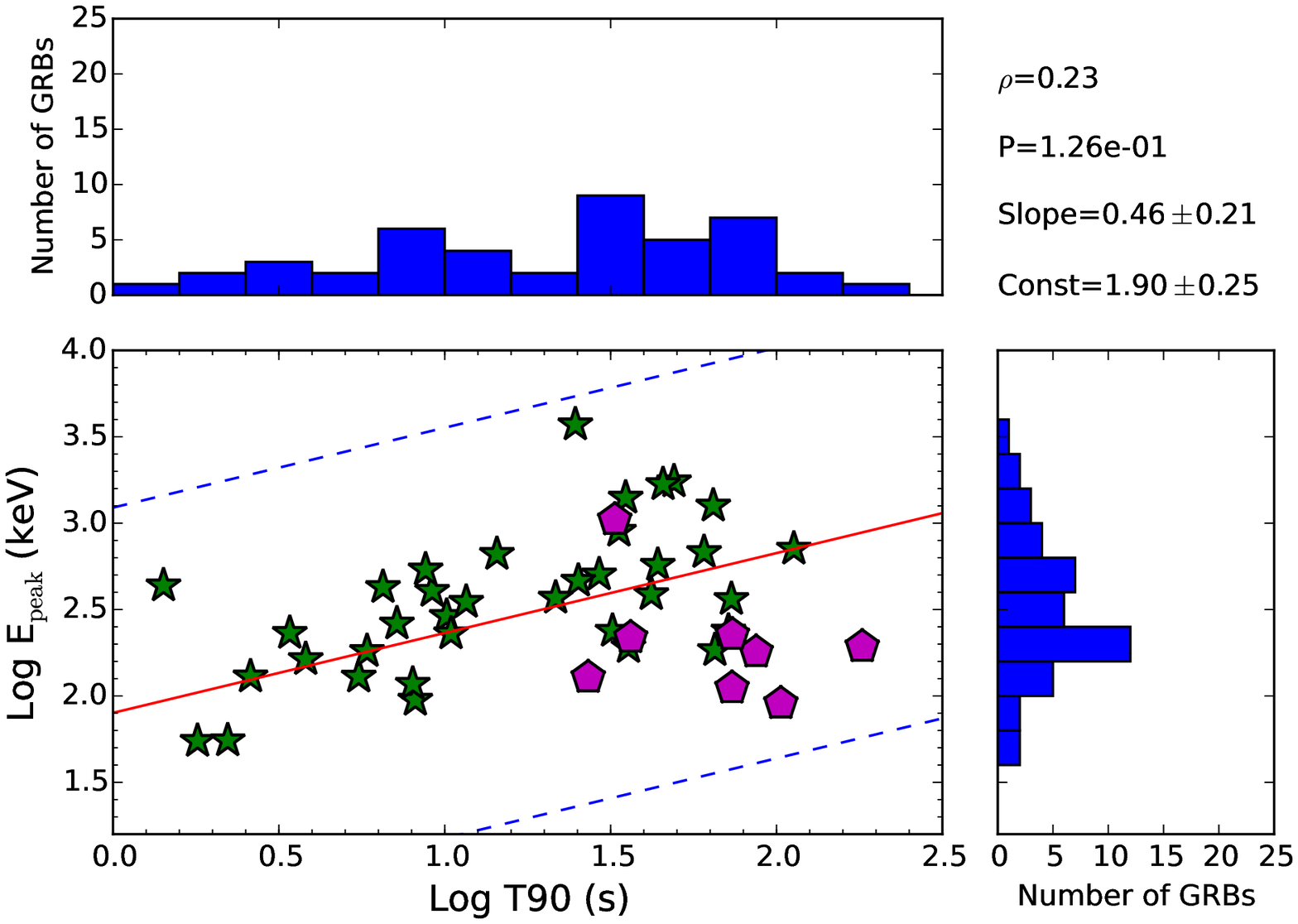}
\caption{In this figure we compare the restframe T90 parameter with the luminosity 
and average decay slopes of the optical/UV and X-ray light curves, and also the isotropic 
energy and peak spectral index. The six panels are Top Left: The optical luminosity
at restframe 200~s versus restframe T90. Top Right: The X-ray luminosity at restframe 200~s versus restframe T90.
Middle Left: The optical average decay index determined from restframe 200~s versus restframe T90. 
Middle Right: The X-ray average decay index determined from restframe 200~s versus restframe T90. 
Bottom Left: The isotropic prompt emission $\rm E_{iso}$ versus restframe T90.
Bottom Right: The $\gamma$-ray peak energy versus restframe T90. 
In all panels, the red solid line represents the best fit regression and the blue dashed line 
represents the 3$\sigma$ deviation. In the top right corner of each panel, we give the Spearman rank coefficient, $\rho$, 
and corresponding null hypothesis probability, P, and we provide the best fit slope 
and constant determined by linear regression. We also overlaid purple pentagons on top of those data points for which the X-ray 
afterglow was contaminated at restframe 200~s by the end of the prompt emission, see \S~.\ref{contam}.}
\label{restT90}
\end{figure*}

In order to ensure that the relationships provided in this paper are not affected by our estimation of
$\rm log\;L_{X,200\rm{s}}$ and $\alpha_{X,>200\rm{s}}$ for the GRBs whose X-ray light curves at 200~s are contaminated by
the prompt emission, we exclude these 8 GRBs and repeat the analysis for all pairs of parameters. We find
that the results do not significantly change for correlations involving all but the restframe T90 parameters.
In these cases, we find that correlations of the different parameters with restframe ${\rm T90}$ provide
Spearman rank coefficients larger than those determined with the same analysis performed on the entire GRB sample.
For most pairs of parameters, the significance implied by the Spearman rank coefficient is $<3\sigma$.
For three pairs of parameters: restframe ${\rm T90}$ and $\rm log\;L_{O,200\rm{s}}$, restframe ${\rm T90}$ and $\rm log\;E_{iso}$,
and restframe ${\rm T90}$ with $\rm log\;E_{peak}$, the significance of correlation implied by the Spearman rank coefficient is $>3\sigma$
and the coefficient suggests strong correlations. However, this is most likely a selection effect. In order to observe the
tail of the prompt emission at restframe 200s, the duration of the prompt emission should be long, but also the tail
of the prompt emission has to be bright enough or the afterglow weak enough so that the emission can be observed above the afterglow.
For these 8 GRBs it is the chance combination of low afterglow luminosity and long duration prompt emission, which
allows the tail of the prompt emission to dominate over the afterglow. It is therefore not a surprise that these
light curves cluster at large restframe ${\rm T90}$ and low $\rm log\;L_{X,200\rm{s}}$. Also since $\rm log\;L_{X,200\rm{s}}$
correlates with $\rm log\;L_{O,200\rm{s}}$ and $\rm log\;E_{iso}$, we should also find clustering when investigating
these parameters with restframe ${\rm T90}$. Examining the corresponding panels of Fig. \ref{restT90}, we find
that the 8 GRBs are clustered in the bottom right of these panels. Therefore by removing these GRBs and
repeating the correlations we are artificially inducing correlations between these parameters.

\section{Discussion}
\label{discussion}
We have explored the restframe properties of a sample of 48 X-ray and optical/UV afterglow light curves.
We have shown that the $\rm log\;L_{200\rm{s}} - \alpha_{>200\rm{s}}$ correlation observed in \cite{oates12}
is also observed in the X-ray light curves. It has been previous suggested that the brightest X-ray
afterglows decay more quickly than fainter afterglows \citep{boe00,kou04,gen05}, which was based
predominantly on pre-{\it Swift} observations of late time X-ray afterglows. However a larger
sample including some of the first {\it Swift} X-ray light curves \citep{gen08} was not able to support previous
claims \citep[see also][]{racusin15}. In this paper, the correlation between X-ray luminosity and temporal behaviour
examines the light curves from a much earlier time, when there is greater spread in the luminosity distribution,
and the average decay index is determined using almost the entire observed afterglow. Since both the optical/UV
and X-ray light curves show $\rm log\;L_{200\rm{s}} - \alpha_{>200\rm{s}}$ correlations, which are consistent, this
points towards a common underlying mechanism producing both the X-ray and optical afterglows. We can therefore
generally exclude models that invoke different emission mechanisms that separately produce the X-ray and optical/UV afterglow.

We also have shown that the X-ray and optical/UV $\rm log\;L_{200\rm{s}}$ are correlated with $\rm log\;E_{iso}$ and $\rm E_{peak}$. This is
consistent with previous studies \citep[e.g.][]{kou04,depas06,nys09,kan10}, particularly with \cite{dav12}
and \cite{mar13}, who performed a similar study using early X-ray luminosity, approximately $5-10$ minutes after trigger.
We have shown for the first time a correlation which relates the average temporal behaviour with $\rm log\;E_{iso}$ and $\rm E_{peak}$. Combined
with the other correlations reported in this paper, this indicates that the GRBs with the brightest, fastest, decaying
afterglows also have the largest observed prompt emission energies and typically larger peak spectral energy.

We now investigate if these observations are consistent with the predictions of the standard synchrotron model 
in its most simple form by comparing our observations with the analytical relationships predicted in \S~\ref{theory}
and with the Monte Carlo simulations described in \S~\ref{MC}. We first verify that the basic properties of our simulation
are consistent with basic properties of the GRB sample. We find the simulation produces the following average values:
$\rm log\;L_{O,200\rm{s}}=32.04\pm 1.26$, $\rm log\;L_{X,200\rm{s}}=30.15\pm 1.02$, $\alpha_{O,>200\rm{s}}=-1.09\pm0.17$,
$\alpha_{X,>200\rm{s}}=-1.25\pm0.18$. These values are consistent at $1\sigma$ with the weighted averages from our
sample: $\rm log\;L_{O,200\rm{s}}=32.11\pm 0.72$, $\rm log\;L_{X,200\rm{s}}=30.46\pm 0.70$, $\alpha_{O,>200\rm{s}}=-0.97\pm0.07$,
$\alpha_{X,>200\rm{s}}=-1.22\pm0.08$. Since for the majority of GRBs in our sample, the X-ray and optical/UV light curves are
consistent with lying in different parts of the spectrum, we have checked that the peak of the distribution of the synchrotron cooling
frequency is consistent with lying in between the optical and X-ray bands. In this case we obtain an average log frequency of $16.55\pm1.19$.
The simulated samples are therefore consistent with being drawn from the same population as our observed data. This suggests
that our initial starting parameters and assumptions were appropriate and therefore we shall continue to compare the results
of the simulation in Table \ref{MCsim} with the observed data in Table \ref{Spear_lin}. 

\subsection{Comparison of observed and predicted correlations}
The standard afterglow model predicts several relationships between $\rm L_{X,200\rm{s}}$ and $\rm L_{O,200\rm{s}}$. Depending on the
spectral regime a GRB may satisfy one of three relationships; see Eq. \ref{lum_eq}. In our observed sample, see Fig. \ref{OX_comp}, we
are only able to observe one overall clustering of points that when fitted with a linear function produces a relationship different to
those predicted in Eq. \ref{lum_eq}. This is not surprising since the number of GRBs in our sample is relatively small and the relationships
are fairly similar. It is therefore important that we compare the observed behaviour with that predicted from the simulations. The Monte
Carlo simulation suggests we should expect a strong linear relationship between $\rm log\;L_{O,200\rm{s}}$ \& $\rm log\;L_{X,200\rm{s}}$ for
a sample of 48 GRBs. The observed linear regression equation is consistent with that simulated at the $1\sigma$ level. The simulation
predicts a non-linear relationship between $\rm log\;L_{O,200\rm{s}}$ \& $\rm log\;L_{X,200\rm{s}}$. It implies that the brighter the
GRB afterglow, the greater the ratio between the X-ray and optical/UV luminosity at 200~s, such that the X-ray luminosity
increases as $\rm L_{X,200\rm{s}}=\rm L_{O,200\rm{s}}^{0.82}$. 

The standard afterglow model \citep{sar98} also predicts several relationships linking the X-ray and optical/UV temporal indices.
The exact closure relation \citep[e.g][]{zhang06,racusin09,gao13} depends on the density structure of external medium and the location
of the observed spectral bands relative to the synchrotron frequencies. These relationships also relate the spectral index to the temporal index,
which enable a more complete picture of the outflow to be formed. In order to obtain information about the outflow producing the afterglow
emission and the medium in to which it explodes, it is preferable to examine both the temporal and spectral parameters of both the
X-ray and optical/UV of each light curve segment, as has already been explored for many GRBs \citep{depas06,gen06,sta08,cur09,sch11,depas13}.
However, we can get an idea of the locations of the X-ray and optical/UV observing bands relative to each other and the structure of
the external medium just by examining the temporal indices of two observed frequencies. The closure relations predict that the 
difference between optical/UV and X-ray decay rate should either be $\Delta \alpha = 0$, if they lie on the same part 
of the synchrotron spectrum, or $|\Delta \alpha| = 0.25 - 0.5$ with a value of 0.25 if the synchrotron cooling 
frequency lies between the X-ray and optical bands and up to 0.5 if energy injection is also considered. The 
$\Delta \alpha$ is expected to be the same whether GRBs are observed on or off-axis \cite[e.g.,][]{mar10}. We have added lines
representing these expected differences to the bottom panel of Fig. \ref{OX_comp}. The best fit regression line lies above, but
close to, the line $\alpha_{X,>200\rm{s}}=\alpha_{O,>200\rm{s}} -0.25$. This implies that  a constant density medium is preferred and the
cooling frequency is likely to lie between the X-ray and optical/UV bands at least for a large number of events. This is consistent
with recent analyses by \cite{ryk09}, \cite{oates11}, \cite{sch11} and \cite{depas13} and supports our choice of assumptions
in \S \ref{theory} \& \ref{MC}. We note that, while the majority of GRBs in our sample are consistent with lying in a constant
density medium, there are a few GRBs that are consistent with lying in a wind-like medium; these are some of the fastest decaying
and therefore the brightest GRBs in the sample. The possibility of the most energetic GRBs having the fastest decaying afterglows
and occurring in wind environments has also been briefly examined by \cite{depas13} and will be examined in more detail in a forthcoming paper
(De Pasquale et al., in prep). We further note that the average decay index is an idealized measure of the afterglow behaviour.
In reality the light curves are likely to consist of one or more temporal segments. However, the closure relations always
predict that if $\nu_m<\nu_o<\nu_c<\nu_x$, then in a wind environment we should typically see the X-ray light
curve decay more quickly than the optical/UV, in a constant density environment it is the other way round. This occurs even if
energy injection is considered. 

The slope of the best fit regression line of $\alpha_{O,>200\rm{s}}$ versus $\alpha_{X,>200\rm{s}}$ is consistent with being unity,
suggesting that the average decay rates of the X-ray and optical/UV light curves are determined by the same mechanism. Comparing
this to the Monte Carlo simulation, we see that the mean Spearman rank coefficient for the simulation is similar
to that determined for the observed data, 32.9 per cent of the simulated sample have Spearman rank coefficients equal to or greater than
that observed, indicating that the observed relationship is fully consistent with that expected from the standard afterglow model. We also
find that the slopes and constant parameters of the observed linear regression for $\alpha_{O,>200\rm{s}}$ versus $\alpha_{X,>200\rm{s}}$ are
consistent within 1$\sigma$ with those simulated. This suggests that the observed relationship between $\alpha_{O,>200\rm{s}}$ and
$\alpha_{X,>200\rm{s}}$ is consistent with the prediction of the standard afterglow model.

We also examined the relationship between $\rm log\;L_{200\rm{s}}$ versus $\alpha_{>200\rm{s}}$. For both the optical/UV and X-ray, we find
the linear regressions give relationships that are consistent at $1\sigma$. This suggests that the same mechanism is producing both correlations.
Comparing the observations with the simulations, we find 0.0 per cent of the 10000 simulations have Spearman rank coefficients more
negative or equal to that observed. Similarly only 1.5 per cent of the simulations have Spearman rank coefficients equal to or more
negative than that observed for the $\rm log\;L_{O,200\rm{s}}$ versus $\alpha_{O,>200\rm{s}}$ correlation. Comparing the linear regression
parameters for the observed and predicted data, we find that the slopes and constant parameters are inconsistent at $\gtrsim 4\sigma$.
Since the average values of the simulated distributions of $\rm log\;L_{200\rm{s}}$ and $\alpha_{>200\rm{s}}$ are consistent with the mean
values of the observed parameter distributions, this indicates that we are simulating GRBs that are representative of our observed sample.
Therefore this implies that correlations as strong as those observed for both the X-ray and optical/UV light curves should not be expected
to be present in our observed sample.

The standard afterglow model also predicts correlations between the isotropic energy $\rm log\;E_{iso}$ with the afterglow
luminosity $\rm log\;L_{200\rm{s}}$, see Eq. \ref{energy_eq}. Since we see a large fraction of GRBs consistent with the
cooling frequency lying in between the X-ray and optical/UV bands (e.g bottom panel of Fig. \ref{OX_comp}), we may expect the
X-ray points to predominately satisfy the second equation and the optical/UV predominately satisfy the first relation. Yet,
the simulation suggests that a single relationship can explain the optical/UV and X-ray correlations between $\rm log\;E_{iso}$
and $\rm log\;L_{200\rm{s}}$ as the simulated slopes are consistent to within $1\sigma$, which is in agreement with the observed sample.
However, we further examined the $\rm log\;L_{200\rm{s}}$ and $\rm log\;E_{iso}$ correlation by directly comparing the slopes of the
simulations and observations. We find them to be inconsistent at $\gtrsim 3\sigma$, with the slope of the observed relationship
being much shallower than that predicted by the simulation.

Spearman rank correlation of the simulated $\rm log\;E_{iso}$ and $\rm log\;L_{200\rm{s}}$ also suggests that we should be
observing weaker correlations in comparison to what we observe, with 0.3 per cent and 0.06 per cent of the simulations having
Spearman rank coefficients equal to or larger than that observed for the optical/UV and X-ray, respectively. This is likely
related to our choice of efficiency. A wide range in efficiency is likely to introduce more scatter in the relationship
between $\rm log\;E_{iso}$ and $\rm log\;L_{200\rm{s}}$. To explore what effect a narrower efficiency would have, we repeated our
simulation with the efficiency parameter fixed at 0.1 and then again at 0.9. In both cases we found the simulated Spearman
rank correlation values were more consistent with those observed, suggesting that the observed sample has a relatively narrow
range in efficiency. However, the slopes of the simulated and observed relationships remain inconsistent at $\gtrsim 3\sigma$
when fixing the efficiency parameter.

In \S~\ref{theory}, we also determined the expected relationship between the ratio $(L_{O}/L_{X})$ and $E_{k}$. We showed that the expected
range in the ratio should lie between 1.05 and 3.16. Comparing these predictions with Fig. \ref{OX_rat}, we see that the observed values
are consistent with this range and therefore consistent with the standard afterglow model. We also note that we do not see evidence for
or against evolution of this ratio with energy as predicted by the second relationship in Eq. \ref{rat_energy_eq}, however we should
not expect to observe a strong correlation because the evolution is very shallow as indicated by the dotted line in Fig. \ref{OX_rat}.

Finally, we also observe that the observed relationships between $\rm log\;E_{iso}$ and $\alpha_{>200\rm{s}}$ for the X-ray and optical/UV
are consistent at 1$\sigma$. We find that only 0.01 per cent of the simulations predict the same or stronger relationship between
$\rm log\;E_{iso}$ and $\alpha_{X,>200\rm{s}}$ and 0.03 per cent of simulations predict similar or stronger relationship between
$\rm log\;E_{iso}$ and $\alpha_{O,>200\rm{s}}$. The slopes and constant parameters for the linear regression from the simulation are
inconsistent with the observed data at $\gtrsim 2.3\sigma$. This suggests that the relationships given in Table \ref{Spear_lin}, for
both the X-ray and optical/UV light curves, between $\rm log\;E_{iso}$ and $\alpha_{>200\rm{s}}$ are not expected in the
standard afterglow model. The lack of strong correlation predicted by the simulation is to be expected since the temporal slopes given
by the closure relations \citep[e.g.,][]{zhang06,racusin09} are a function of the electron energy index only and are not seemingly
directly related to the energy of the outflow.

Overall we would expect to see relationships observed between $\rm log\;L_{O,200\rm{s}}$ \& $\rm log\;L_{X,200\rm{s}}$ and 
$\alpha_{O,>200\rm{s}}$ versus $\alpha_{X,>200\rm{s}}$ arise because the same afterglow is observed in both the X-ray and optical/UV. These 
relationships can be explained easily by the standard afterglow model and are fully consistent with our simulations. Also a relationship 
between $\rm log\;E_{iso}$ versus $\rm log\;L_{200\rm{s}}$ is expected in the standard afterglow model, however, comparison of our observed
relationship to the simulations suggests that the observed linear regression slope is less steep than predicted by the simulation.
Furthermore, the relationships we observe, between $\rm log\;L_{200\rm{s}}$ and $\alpha_{>200\rm{s}}$, and $\rm log\;E_{iso}$ and
$\alpha_{>200\rm{s}}$, are not predicted by the simulations and are not expected in the standard afterglow model. Since the standard
afterglow does not succeed in fully predicting all of our observed correlations, it is therefore likely that a more complex outflow
model is required. This conclusion is similar to that drawn during the separate investigation of the optical/UV
$\rm log\;L_{200\rm{s}}-\alpha_{>200\rm{s}}$ decay correlation. 

To summarize, we find that the optical/UV and X-ray afterglows are strongly related and it is likely that 
they are produced by the same outflow and by the same or at least related mechanisms. However, as indicated above the basic standard 
afterglow model does not predict all of our observed correlations and it is therefore likely that a more complex outflow model is 
required to explain all the observed correlations. 

\subsection{Alternative Models}
There are two main possibilities that could make the outflow complex enough to be able to reproduced the observed correlations. The 
first is that perhaps there is some mechanism or parameter that controls the amount of energy given to and distributed during the prompt 
and afterglow phases and that also regulates the afterglow decay rate. This should occur in such a way that for events with the largest
gamma-ray isotropic energy, the energy given to the afterglow is released quickly, resulting in an initially bright afterglow which 
decays rapidly. Conversely, if the gamma-ray isotropic energy is smaller, then the afterglow energy is released slowly over a longer 
period, the afterglow will be less bright initially and decay at a slower rate.

The second possibility is that the correlations could be a geometric effect, perhaps the result of the observer's viewing angle. Jets 
viewed away from the jet-axis may have fainter afterglows that decay less quickly in comparison to afterglows viewed closer to the center 
of the jet \citep[see Fig 3. of][]{pan08}. Similarly, this will also affect the observed prompt emission, with jets viewed off-axis appearing to have 
lower isotropic energy and lower peak spectral energy \citep{ram05}. In this case, the relationship between luminosity and decay rate of GRB afterglow
light curves should be observed in uniform jets and in structured outflows \citep{pan08}. By looking at Figure 3 of \cite{pan08},
two further tests can be derived to determine if this scenario is producing the observed afterglows. The first is that we should expect to
see convergence of the light curves at late times to a similar decay rate for all observing angles when the emission from the entire jet
is observed by the observer. The convergence time and the range of decay rates will vary, depending on how the outflow is structured. The
second is that afterglows that are viewed more off-axis will rise later. Therefore we should also observe a correlation between afterglow brightness
and peak time, although the strength of this correlation will be affected by whether or not GRBs have similar jet structure. This latter
test has been explored by \cite{pan08}, \cite{pan11} and \cite{pan13}. They find a significant correlation between the peak time
and peak afterglow brightness in both the X-ray and optical light curves consistent with this hypothesis. However, we note that this
correlation was determined from GRBs with observed rises and therefore afterglows that peak before observations begin will not have
been included \citep{pan13}.

%\vspace{-1mm}
\section{Conclusions}
\label{conclusions}
In the optical/UV GRB afterglow sample of \cite{oates12} a correlation was found between the early optical/UV luminosity
(measured at restframe 200~s, $\rm log\;L_{200\rm{s}}$) and average decay rate (measured from 200~s, $\alpha_{>200\rm{s}}$).
The aim of this paper was to explore whether this was also observed in the X-ray light curves, to explore how this correlation relates to the
prompt emission phase and to explore if what we see is consistent with the predictions of the standard afterglow model. 

We first began by exploring what relationships the standard afterglow model predicts for our observed parameters. For different ordering
of the spectral frequencies, this model predicts more than one expression for the relationship between two parameters. It is therefore
not possible to analytically predict the expected correlations for a sample of GRBs with a mixture of spectral regimes. Therefore, we
performed a Monte Carlo simulation to predict the relationships between various combinations of parameters for a sample of 48 GRBs.

We then examined the afterglow parameters and correlations resulting from the observed sample and compared them to the prediction of the simulations.
We find luminosity-decay correlations in both the optical/UV and X-ray light curves and find that these relationships are consistent
\citep[see also][]{racusin15}. This suggests a single underlying mechanism producing the correlations in both bands, regardless of
their detailed temporal behaviour. We also show significant correlations between the logarithmic X-ray and optical/UV luminosity
($\rm log\;L_{O,200\rm{s}}$, $\rm log\;L_{X,200\rm{s}}$) and the optical/UV and X-ray decay indices ($\alpha_{O,>200\rm{s}}$ and
$\alpha_{X,>200\rm{s}}$). These relationships are predicted by the standard afterglow model and the observations are consistent
with our simulations. However such strong correlations between $\rm log\;L_{200\rm{s}}$ and $\alpha_{>200\rm{s}}$ at both wavelengths
are not predicted in the standard afterglow model and are inconsistent with our simulations.

Finally we compared the parameters in the both the X-ray and optical/UV luminosity-decay correlations with the prompt emission parameters,
such as isotropic energy ($\rm E_{\rm iso}$), restframe peak spectral energy ($\rm E_{peak}$) and the restframe T90 parameter
(duration over which 90 per cent of the emission is observed). We show significant evidence that the X-ray and optical 
luminosities, measured at 200~s, are correlated with $\rm E_{\rm iso}$ and slightly less strongly correlated with $\rm E_{peak}$. This
is consistent with previous findings \citep[e.g.][]{dav12,mar13} and predictions of the standard afterglow model and our simulations,
although the slopes of the relationships between luminosity and isotropic energy are steeper in the simulations than observed. The
average decay indices for the X-ray and optical/UV bands are also correlated with $\rm E_{\rm iso}$, but these correlations are slightly
weaker in comparison with those correlations observed for the luminosities at 200~s and $\rm E_{\rm iso}$. The observed relationships
between $\alpha_{>200\rm{s}}$ and $\rm E_{\rm iso}$ are not expected in the standard afterglow model and are inconsistent with our simulations.

Together these correlations imply that the GRBs with the brightest afterglows in the X-ray and optical bands, decay the fastest
and they also have the largest observed prompt emission energies and typically larger peak spectral energy. This suggests that
what happens during the prompt phase has direct implications on the afterglow.

Overall, while correlations between the luminosities in both the X-ray and optical/UV bands, between the decay indices and between
the luminosities and the isotropic energy are predicted by the simulation of the standard afterglow model, observed relationships
involving the average decay indices with either luminosity at 200~s or the isotropic energy are not consistent with the standard afterglow model. 
We therefore suggest that a more complex afterglow or outflow model is required to produce all the observed correlations. This may be 
due to either a viewing angle effect or by some mechanism or physical property controlling the energy release within the outflow.

\section{Acknowledgments}
We thank the referee for providing critical comments and suggestions that have helped to improve this paper. We also thank Amy Lien
for providing {\it Swift} BAT parameters for GRB~071112C. This research has made use of data obtained from the High Energy Astrophysics
Science Archive Research Center (HEASARC) and the UK Swift Science Data Centre provided by NASA's Goddard Space Flight Center and
the University of Leicester, UK, respectively. SRO, MDP, MJP, AAB, NPMK and PJS acknowledge the support of the UK Space Agency.
SRO also acknowledges the support of the Spanish Ministry, Project Number AYA2012-39727-C03-01.         
%                       
                       
%%%%%%%%%%%%%%%%%%%%%%%%%%%%%%%%%%%%%%%%%%%%
                       
\bibliographystyle{mn2e}   % if natbib is available
\bibliography{LUM_DECAY_CORR} %bibtex file
                       
\IfFileExists{\jobname.bbl}{}
 {\typeout{}           
  \typeout{******************************************}
  \typeout{** Please run "bibtex \jobname" to optain}
  \typeout{** the bibliography and then re-run LaTeX}
  \typeout{** twice to fix the references!}
  \typeout{******************************************}
  \typeout{}           
 }                     

\newpage
\section*{Appendix}    
                       
\setcounter{table}{0}
\renewcommand{\thetable}{A\arabic{table}}
\begin{table*}
  \begin{tabular}{lcccccccc}
    \hline
    \footnotesize
    GRB & Redshift &  $\rm log\;L_{O,200\rm{s}}$&  $\rm log\;L_{X,200\rm{s}}$ & $\alpha_{O,>200\rm{s}}$  & $\alpha_{X,>200\rm{s}}$ & $\rm log\;E_{\rm iso}$ & $\rm E_{peak}$ & T90\\
    \hline
    GRB050319    &  $3.2425^{1}$    &  $31.84\pm0.26$  &  $29.76\pm0.08$  &  $-0.68\pm0.06$  &  $-0.84\pm0.02$  & 52.71   & 45    & 151.7 \\  
    GRB050525A   &  $0.606^{2}$     &  $31.30\pm0.04$  &  $29.82\pm0.09$  &  $-1.10\pm0.01$  &  $-1.53\pm0.04$  & 52.42   & $80^{a}$& 8.8   \\
    GRB050730    &  $3.9693^{3,4}$  &  $32.75\pm0.08$  &  $31.16\pm0.02$  &  $-1.31\pm0.11$  &  $-2.00\pm0.02$  & 53.14   & 101   & 145.1 \\
    GRB050801    &  $1.38^{3,5}$    &  $31.68\pm0.02$  &  $29.29\pm0.11$  &  $-1.28\pm0.10$  &  $-1.15\pm0.04$  & 51.42   & 40    & 19.4  \\
    GRB050802    &  $1.71^{6}$      &  $31.89\pm0.09$  &  $30.24\pm0.06$  &  $-0.80\pm0.03$  &  $-1.18\pm0.01$  & 52.51   & 107   & 27.5  \\
    GRB050922C   &  $2.1995^{1}$    &  $32.40\pm0.05$  &  $30.52\pm0.03$  &  $-1.03\pm0.03$  &  $-1.30\pm0.02$  & 52.60   & 136   & 4.5   \\
    GRB060418    &  $1.49^{7}$      &  $32.60\pm0.01$  &  $30.50\pm0.02$  &  $-1.24\pm0.01$  &  $-1.40\pm0.02$  & 53.14   & $230^{b}$ & 109.2 \\
    GRB060510A   &  $1.2^{8}$       &  $32.11\pm0.05$  &  $30.08\pm0.04$  &  $-0.48\pm0.05$  &  $-1.01\pm0.01$  & 52.89   & $184^{c}$ & 20.2  \\   
    GRB060512    &  $2.100^{3,9}$   &  $31.70\pm0.06$  &  $30.13\pm0.09$  &  $-1.05\pm0.05$  &  $-1.17\pm0.05$  & 51.86   & --    & 11.4  \\ 
    GRB060526    &  $3.2213^{1}$    &  $32.26\pm0.10$  &  $29.94\pm0.09$  &  $-0.87\pm0.06$  &  $-0.99\pm0.03$  & 52.67   & 44    & 275.2 \\
    GRB060605    &  $3.773^{10}$    &  $32.91\pm0.12$  &  $30.27\pm0.08$  &  $-1.15\pm0.11$  &  $-1.57\pm0.04$  & 52.86   & 148   & 539.1 \\
    GRB060607A   &  $3.0749^{11}$   &  $32.62\pm0.05$  &  $30.79\pm0.03$  &  $-1.20\pm0.06$  &  $-1.59\pm0.02$  & 53.01   & 114   & 103.0 \\
    GRB060708    &  $1.92^{5}$      &  $31.50\pm0.14$  &  $29.74\pm0.05$  &  $-0.82\pm0.35$  &  $-1.17\pm0.02$  & 51.92   & 79    & 10.0  \\
    GRB060729    &  $0.5428^{3,12}$ &  $30.23\pm0.08$  &  $28.74\pm0.04$  &  $-0.63\pm0.01$  &  $-0.82\pm0.01$  & 51.55   & 72     & 113.0 \\
    GRB060908    &  $1.8836^{13}$   &  $31.53\pm0.10$  &  $30.28\pm0.07$  &  $-1.14\pm0.13$  &  $-1.46\pm0.05$  & 52.74   & $148^{a}$& 18.8  \\
    GRB060912A   &  $0.937^{14}$    &  $31.04\pm0.16$  &  $29.00\pm0.06$  &  $-0.55\pm0.05$  &  $-1.10\pm0.03$  & 51.75   & 67     & 5.0   \\
    GRB061007    &  $1.262^{15}$    &  $33.28\pm0.02$  &  $31.05\pm0.01$  &  $-1.60\pm0.02$  &  $-1.68\pm0.01$  & 53.91   & $399^{d}$  & 75.7  \\ 
    GRB061021    &  $0.346^{3}$     &  $30.27\pm0.04$  &  $28.50\pm0.06$  &  $-0.86\pm0.01$  &  $-0.95\pm0.01$  & 51.82   & $778^{e}$  & 43.8  \\ 
    GRB061121    &  $1.314^{16}$    &  $31.88\pm0.09$  &  $30.10\pm0.02$  &  $-0.70\pm0.02$  &  $-1.09\pm0.01$  & 53.55   & $606^{f}$ & 81.2  \\
    GRB070318    &  $0.8397^{3,17}$ &  $31.97\pm0.05$  &  $29.53\pm0.03$  &  $-1.00\pm0.02$  &  $-1.17\pm0.02$  & 52.02   & 130    & 131.5 \\
    GRB071112C   &  $0.823^{18}$    &  $31.00\pm0.06$  &  $29.48\pm0.03$  &  $-1.23\pm0.07$  &  $-1.45\pm0.02$  &  --     & --     & 44.80 \\  
    GRB080310    &  $2.4274^{3,19}$ &  $31.90\pm0.11$  &  $29.39\pm0.08$  &  $-0.89\pm0.04$  &  $-1.13\pm0.03$  & 52.76   & 26     & 352.4 \\
    GRB080319B   &  $0.9382^{3,20}$ &  $33.01\pm0.03$  &  $31.49\pm0.01$  &  $-1.48\pm0.01$  &  $-1.64\pm0.01$  & 54.07   & $651^{g}$   & 124.9 \\ 
    GRB080413B   &  $1.1014^{3,21}$ &  $31.67\pm0.14$  &  $29.88\pm0.04$  &  $-0.73\pm0.04$  &  $-1.02\pm0.01$  & 52.27   & $78^{a}$  & 8.0   \\
    GRB080430    &  $0.767^{22}$    &  $30.46\pm0.08$  &  $28.73\pm0.04$  &  $-0.70\pm0.03$  &  $-0.78\pm0.01$  & 51.51   & 66     & 14.2  \\
    GRB080721    &  $2.5914^{3,23}$ &  $32.93\pm0.09$  &  $31.61\pm0.01$  &  $-1.14\pm0.05$  &  $-1.40\pm0.01$  & 54.01   & $485^{h}$  & 176.3 \\
    GRB080804    &  $2.2045^{24}$   &  $31.67\pm0.04$  &  $29.80\pm0.06$  &  $-0.94\pm0.03$  &  $-1.11\pm0.02$  & 52.94   & $109^{i,j}$ & 37.2  \\
    GRB080810    &  $3.3604^{3}$    &  $33.34\pm0.10$  &  $30.61\pm0.03$  &  $-1.16\pm0.09$  &  $-1.58\pm0.03$  & 53.84   & $856^{i,j}$ & 107.7 \\
    GRB080916A   &  $0.689^{25}$    &  $31.08\pm0.19$  &  $28.97\pm0.06$  &  $-0.79\pm0.06$  &  $-0.97\pm0.02$  & 52.05   & $129^{a}$& 61.3  \\
    GRB080928    &  $1.6919^{3,26}$ &  $31.06\pm0.19$  &  $30.05\pm0.05$  &  $-1.29\pm0.06$  &  $-1.55\pm0.04$  & 52.49   & 67     & 233.7 \\
    GRB081007    &  $0.5295^{27}$   &  $30.20\pm0.06$  &  $28.69\pm0.09$  &  $-0.70\pm0.02$  &  $-0.88\pm0.01$  & 51.06   & --     & 9.0   \\
    GRB081008    &  $1.967^{28}$    &  $32.82\pm0.08$  &  $30.10\pm0.02$  &  $-1.09\pm0.02$  &  $-1.25\pm0.02$  & 53.06   & $229^{i,j}$ & 179.5 \\
    GRB081203A   &  $2.100^{29}$    &  $33.55\pm0.02$  &  $30.60\pm0.02$  &  $-1.52\pm0.01$  &  $-1.43\pm0.02$  & 53.21   & 119    & 223.0 \\
    GRB081222    &  $2.77^{30}$     &  $32.46\pm0.04$  &  $30.93\pm0.02$  &  $-0.93\pm0.03$  &  $-1.22\pm0.01$  & 53.27   & $143^{i,j}$ & 33.0  \\
    GRB090401B   &  $3.1^{5}$       &  $32.49\pm0.01$  &  $31.37\pm0.01$  &  $-1.71\pm0.16$  &  $-1.45\pm0.01$  & 54.01   & $409^{k}$ & 186.5 \\ 
    GRB090418A   &  $1.608^{31}$    &  $33.13\pm0.15$  &  $30.29\pm0.07$  &  $-1.18\pm0.22$  &  $-1.27\pm0.02$  & 52.83   & 142    & 56.3  \\
    GRB090424    &  $0.544^{32}$    &  $31.64\pm0.07$  &  $30.12\pm0.01$  &  $-0.75\pm0.02$  &  $-1.11\pm0.01$  & 52.56   & $154^{a}$& 49.5  \\
    GRB090618    &  $0.54^{33}$     &  $31.59\pm0.01$  &  $30.09\pm0.01$  &  $-0.99\pm0.01$  &  $-1.41\pm0.01$  & 53.41   & $147^{i,j}$ & 113.3 \\
    GRB090812    &  $2.452^{34}$    &  $32.42\pm0.13$  &  $30.69\pm0.02$  &  $-1.19\pm0.35$  &  $-1.16\pm0.02$  & 53.97   & --    & 75.1  \\ 
    GRB091018    &  $0.971^{35}$    &  $31.69\pm0.02$  &  $29.85\pm0.03$  &  $-0.96\pm0.01$  &  $-1.18\pm0.01$  & 51.79   & $28^{l}$  & 4.4   \\
    GRB091020    &  $1.71^{36}$     &  $33.06\pm0.10$  &  $30.35\pm0.03$  &  $-1.17\pm0.06$  &  $-1.20\pm0.01$  & 52.92   & $244^{i,j}$ & 38.9  \\
    GRB091029    &  $2.752^{37}$    &  $31.54\pm0.11$  &  $30.01\pm0.05$  &  $-0.72\pm0.02$  &  $-0.81\pm0.01$  & 52.85   & $61^{a}$ & 39.2  \\
    GRB091208B   &  $1.063^{38}$    &  $31.69\pm0.21$  &  $29.69\pm0.06$  &  $-0.79\pm0.09$  &  $-1.06\pm0.02$  & 52.31   & $127^{i,j}$ & 14.8  \\
    GRB100316B   &  $1.180^{39}$    &  $31.31\pm0.12$  &  $29.07\pm0.13$  &  $-0.94\pm0.09$  &  $-1.09\pm0.06$  & 51.12   & 25     & 3.9   \\
    GRB100805A   &  $1.85^{5}$      &  $32.11\pm0.11$  &  $29.40\pm0.08$  &  $-0.73\pm0.12$  &  $-1.63\pm0.02$  & 51.88   & 64     & 16.6  \\
    GRB100901A   &  $1.408^{40}$    &  $31.36\pm0.05$  &  $28.07\pm0.09$  &  $-0.62\pm0.01$  &  $-1.02\pm0.02$  & 52.26   & 80     & 436.4 \\
    GRB100906A   &  $1.727^{41}$    &  $32.83\pm0.02$  &  $30.06\pm0.03$  &  $-1.13\pm0.01$  &  $-1.30\pm0.02$  & 53.57   & $142^{m}$ & 114.3 \\
    GRB101219B   &  $0.5519^{42}$   &  $30.56\pm0.08$  &  $27.47\pm0.13$  &  $-0.82\pm0.04$  &  $-0.65\pm0.03$  & 51.59   & $83^{i,j}$ & 42.0  \\
    %[-2.0ex]              
    \hline                 
    \hline
  \end{tabular}

  \caption{Table containing all the parameters for all the GRBs in the sample: X-ray and optical luminosity at restframe 200~s $\rm log\;L_{200\rm{s}}$, 
    average decay indices of the X-ray and optical/UV light curves measured using data from restframe 200~s, $\alpha_{>200\rm{s}}$; and observer frame 
    values for isotropic energy $\rm E_{\rm iso}$, gamma-ray peak energy $\rm E_{peak}$, and duration of the prompt emission. For $\rm E_{peak}$, where 
    no reference is given we used the correlation between the peak energy and the photon index of the $\nu F_\nu$ spectrum to estimate $\rm E_{\rm peak}$ 
    \citep[see][for further details]{sak09}. The relationship can only be used to estimate $\rm E_{\rm peak}$ when the power-law index of the BAT spectrum 
    is between -2.3 and -1.3, which places $\rm E_{peak}$ approximately within the BAT range. References: 1) \protect\cite{jak06},2) \protect\cite{3483}, 3)\protect\cite{fyn09}, 
    4)\protect\cite{3746}, 5)\protect\cite{oates12}, 6)\protect\cite{3749}, 7)\protect\cite{4974}, 8)\protect\cite{oates12}, 9)\protect\cite{5131}, 10)\protect\cite{fer09}, 11)\protect\cite{fox08}, 12)\protect\cite{5373}, 
    13)\protect\cite{5555}, 14)\protect\cite{5617}, 15)\protect\cite{5716}, 16)\protect\cite{5826}, 17)\protect\cite{6216}, 18)\protect\cite{7076}, 19)\protect\cite{7388}, 20)\protect\cite{7444}, 21)\protect\cite{7601}, 
    22)\protect\cite{7654}, 23)\protect\cite{7998}, 24)\protect\cite{8058}, 25)\protect\cite{8254}, 26)\protect\cite{8301}, 27)\protect\cite{8335}, 28)\protect\cite{8346}, 29)\protect\cite{kui09}, 30)\protect\cite{8713}, 
    31)\protect\cite{9151}, 32)\protect\cite{9243}, 33)\protect\cite{9518}, 34)\protect\cite{9771}, 35)\protect\cite{10038}, 36)\protect\cite{10053}, 37)\protect\cite{10100}, 38)\protect\cite{10263}, 39)\protect\cite{10495}, 
    40)\protect\cite{11164}, 41)\protect\cite{11230}, 42)\protect\cite{11579}.  a)\protect\cite{sak11}, b)\protect\cite{4989}, c)\protect\cite{5113}, d)\protect\cite{5722}, e)\protect\cite{5748}, f)\protect\cite{5837}, 
    g)\protect\cite{7482}, h)\protect\cite{7995}, i)\protect\cite{gol12}, j)\protect\cite{gru14}, k)\protect\cite{9083}, l)\protect\cite{10045}, m)\protect\cite{11251}}
  \label{appenx1}
\end{table*}

\end{document}